%% file: main.tex
\documentclass[camera,letterpaper,nomarginnotes,nonarrowgutter]{jpaper}
\usepackage[numbers,sort&compress,square]{natbib}
\usepackage{amsmath,amssymb,amsfonts}
\usepackage{algorithmic}
\usepackage{graphicx}
\usepackage{textcomp}
\usepackage{xcolor}
\usepackage{fancyhdr}
\usepackage{xspace}
\usepackage{booktabs}
\usepackage{enumitem}
\usepackage{etoolbox}

\usepackage[colorlinks=true]{hyperref}

\usepackage{tikz}
\usepackage{soul}
\usepackage{multirow}
\usepackage{graphicx}
\usepackage{array}
\newcolumntype{C}[1]{>{\centering\let\newline\\\arraybackslash\hspace{0pt}}m{#1}}
            
\def\BibTeX{{\rm B\kern-.05em{\sc i\kern-.025em b}\kern-.08em
    T\kern-.1667em\lower.7ex\hbox{E}\kern-.125emX}}

\newcommand*\circledb[1]{\tikz[baseline=(char.base)]{
            \node[shape=circle,fill,inner sep=0.5pt] (char) {\textcolor{white}{#1}};}}
\newcommand*\circleda[1]{\tikz[baseline=(char.base)]{
            \node[shape=circle,fill,inner sep=1.2pt] (char) {\textcolor{white}{#1}};}}            

\fancypagestyle{firstpage}{
  \fancyhf{}

  \fancyfoot[C]{\thepage}
}
\newcommand{\techL}{AgileWatts\xspace}
\newcommand{\tech}{AW\xspace}
\newcommand{\CAgile}{C6A\xspace}

\newcommand\hj[1]{{\color{black}{#1}}}
\newcommand\jk[1]{{\color{black}{#1}}}
\newcommand\hjj[1]{{\color{black}{#1}}}

\newcommand\hjjj[1]{{\color{black}{#1}}}

\newcommand{\hbb}[2][cyan]{{%
     #2}%
}

\newcommand{\hgg}[2][green]{{%
 #2}%
}

\newcommand{\hyy}[2][yellow]{{%
    #2}%
}
\newcommand{\hpp}[2][pink]{{%
    #2}%
}
\newcommand{\hoo}[2][orange]{{%
    #2}%
}
\newcommand{\hy}[2][yellow]{{%
    #2}%
}
\newcommand{\hp}[2][pink]{{%
    #2}%
}
\newcommand{\ho}[2][orange]{{%
    #2}%
}

 \newcommand{\hg}[2][green]{{%
    #2}%
 }

 \newcommand{\hb}[2][cyan]{{%
    #2}%
}

\newcommand{\rG}[1]{%
{\color{black}#1}%
}

 \title{  \textit{\techL}: An Energy-Efficient CPU Core Idle-State \\ Architecture for Latency-Sensitive Server Applications}
\author{\vspace{-10pt}\\
{Jawad Haj Yahya$^{1}$} \quad \quad
{Haris Volos$^{2}$} \quad \quad
{Davide B. Bartolini$^{1}$} \quad \quad
{Georgia Antoniou$^{2}$} 
\vspace{5pt}\\
{Jeremie S. Kim$^{3}$} \quad \quad
{Zhe Wang$^{1}$} \quad \quad
{Kleovoulos Kalaitzidis$^{1}$} \quad \quad
{Tom Rollet$^{1}$} 
\vspace{5pt}\\
{Zhirui Chen$^{1}$} \quad \quad
{Ye Geng$^{1}$} \quad \quad
{Onur Mutlu$^{3}$} \quad \quad
{Yiannakis Sazeides$^{2}$} 
\vspace{12pt}\\
{\fontsize{10}{11}\selectfont
$^{1}$\textit{Huawei Technologies}\quad \quad
$^{2}$\textit{University of Cyprus} \quad \quad
$^{3}$\textit{ETH Zurich}
}
\vspace{0pt}}

\begin{document}
\bstctlcite{IEEEexample:BSTcontrol}
\maketitle
\thispagestyle{plain}
\pagestyle{plain}


\AtEndEnvironment{thebibliography}{
\bibitem{B1}  Barroso, Luiz André, Jeffrey Dean, and Urs Holzle. ``Web search for a planet: The Google cluster architecture.'' IEEE Micro 2003. 
\bibitem{B2} Jeon, M., He, Y., Elnikety, S., Cox, A. and Rixner, S. ``Adaptive parallelism for web search''. EuroSys 2013.
\bibitem{B3} D. Wong and M. Annavaram. ``Knightshift: Scaling the Energy Proportionality Wall Through Server-level
Heterogeneity''. In MICRO, 2012.
\bibitem{B4} S. Luo, H. Xu, C. Lu, K. Ye, G. Xu, L. Zhang, Y. Ding, J. He, and C. Xu. ``Characterizing Microservice Dependency and Performance: Alibaba Trace Analysis''. In SoCC, 2021.
\bibitem{mysql} Oracle, ``MySQL Workbench'', online, accessed August 2022, \url{https://www.mysql.com/products/workbench/} .
\bibitem{spark} Github, ``Spark-Bench'', online, accessed August 2022, \url{https://codait.github.io/spark-bench/}
\bibitem{hive} Apache , ``Apache Hive'', online, accessed August 2022, \url{https://hive.apache.org/} 
\bibitem{B5} Jiang, Hailin, Malgorzata Marek-Sadowska, and Sani R. Nassif. ``Benefits and Costs of Power-gating Technique.'' ICCD, 2005.
\bibitem{B6} James Lewis and Martin Fowler. ``Microservices.'' online, accessed June 2022 \url{https://martinfowler.com/articles/microservices.html}
\bibitem{B7} Twitter. ``Decomposing Twitter: Adventures in Service Oriented Architecture.'' online, accessed June 2022 \url{https://www.slideshare.net/InfoQ/decomposing-twitter-adventures-in-serviceoriented-architecture}
\bibitem{B8} Rob Brigham. ``DevOps at Amazon: A Look at Our Tools and Processes.'' online, accessed June 2022 \url{https://www.slideshare.net/AmazonWebServices/devops-at-amazon-a-look-at-our-tools-and-processes}
\bibitem{Jalili2021} Jalili, Majid, et al. ``Cost-efficient overclocking in immersion-cooled datacenters.'' ISCA 2021.
\bibitem{EnergyPrice} Global Petrol Prices, Electricity prices for households, September 2021 \url{https://www.globalpetrolprices.com/electricity_prices/}
\bibitem{E1} Schöne, Robert, et al. "Energy efficiency aspects of the AMD Zen 2 architecture." CLUSTER 2021. 
\bibitem{E2} AMD EPYC 7313P Energy Consumption Test, \url{https://metebalci.com/blog/epyc-energy-consumption-test/} 
\bibitem{E3} Tuning UEFI Settings for Performance and Energy Efficiency on AMD Processor-Based ThinkSystem Servers  \url{https://lenovopress.lenovo.com/lp1267.pdf} 
\bibitem{E4} Performance Tuning for Cisco UCS C225 M6 and C245 M6 Rack Servers with 3rd Gen AMD EPYC \url{https://www.cisco.com/c/en/us/products/collateral/servers-unified-computing/ucs-c-series-rack-servers/performance-tuning-wp.html} 

}

\newcommand\fixme[1]{{\color{red} {\bf FIXME:} #1}}
\newcommand\fixremove[1]{{\color{purple} {\bf REMOVE:} #1}}
\newcommand\fixadd[1]{{\color{blue} {\bf ADD:} #1}}

\begingroup\renewcommand\thefootnote{$\star$}
\footnotetext{\,The first and fourth authors performed this work while affiliated with the Computing Systems Lab, Huawei Zurich Research Center.}

\endgroup
\input{_00_abstract}

\input{_01_introduction}

\input{_02_motivation}
\input{_03_background}
\input{_04_technique}

\input{_05_implementation}

\input{_06_methodology}
\input{_07_results}

\input{_08_relatedwork}

\input{_09_conclusion}

\section*{Acknowledgments}
\begin{sloppypar}
We thank the anonymous reviewers of ISCA 2022 and MICRO 2022 for their constructive critique and feedback. \jk{We thank the SAFARI Research Group members for valuable feedback and the stimulating intellectual environment they provide.}
This project has received funding from the European Union’s Horizon 2020 research and innovation programme under the Marie Skłodowska-Curie grant agreement No 101029391. 
\end{sloppypar}



\bibliographystyle{IEEEtran}
\bibliography{refs}

\end{document}

%% file: _00_abstract.tex
\begin{abstract} 
User-facing applications running in modern datacenters exhibit irregular request patterns and are implemented using a multitude of services with tight latency requirements ($30$--$250{\mu}s$). These characteristics render existing energy-conserving techniques \jk{ineffective} when processors are idle due to the long transition time (order of $100{\mu}s$) from a deep CPU core idle power state (C-state). While prior works propose management  techniques to mitigate this inefficiency, we tackle it at its root with \techL (\tech): a new deep CPU core C-state architecture optimized for datacenter server processors targeting latency-sensitive applications.

\tech drastically reduces the transition latency \jk{from} deep CPU core idle power states while retaining most of their power savings
based on three key ideas.
First, \tech eliminates the latency 
(several microseconds) of saving/restoring the core context 
when powering-off/-on the core in a deep idle 
state by i) implementing medium-grained power-gates, carefully distributed across the CPU core, and ii) retaining context in the power-ungated domain.
Second, \tech eliminates the flush latency 
(several tens of microseconds) of the L1/L2 caches when entering a deep idle 
state by keeping L1/L2 
content power-ungated. A \jk{small} control logic also remains ungated to serve cache coherence traffic.
\tech implements cache sleep-mode and leakage reduction \jk{for} the power-ungated domain by lowering a core's voltage to the minimum operational level.
Third, using a state-of-the-art power efficient all-digital phase-locked loop (ADPLL) clock generator, \tech keeps the PLL active and locked during the idle state, 
cutting 
microseconds of wake-up latency \jk{at negligible} power cost.


Our evaluation with an accurate \jk{industrial-grade} simulator calibrated against an Intel Skylake server shows  that \tech reduces the energy consumption of Memcached by up to $71\%$ ($35\%$ on average) with \hjj{${<}1\%$  end-to-end} performance degradation. \jk{We observe similar} trends for other evaluated services (MySQL and Kafka). {\tech}'s new deep C-states $C6A$ and $C6AE$ reduce \hj{transition-time} by up to $900\times$ as compared to 
the deepest existing idle state $C6$, while consuming only $7\%$ and $5\%$ of the active state (C0) power, respectively. 
\end{abstract}

%% file: _01_introduction.tex
\section{Introduction}
\label{sec:intro}

\jk{Large datacenters running user facing applications are using inefficiently their servers} \emph{due to the killer microseconds} ~\cite{barroso2017attack,barroso2018datacenter,prekas2017zygos}.
Killer microseconds refer to microsecond-scale \emph{idleness} during CPU execution caused \jk{by a} combination of two major trends.
First, various events (e.g., related to NVM storage, faster datacenter networking, and main memory) with latencies in the range of microseconds are prevalent~\cite{barroso2017attack,chou2019mudpm,cho2018taming}. 
Second, \ho{a new} software architecture is deployed in datacenters based on \emph{microservices}, i.e., a large application composed of numerous interconnected smaller services that explicitly communicate with each other.
\jk{Such} applications exhibit irregular request streams, and their services have very tight (i.e., few tens to few hundreds of microseconds) latency requirements~\cite{dmitry2014micro}.



\autoref{tab:c-states} reports a typical hierarchy of \hjj{\emph{core idle states}} (i.e., \emph{C-states}: $C0$, $C1$, $C1E$, and $C6$)\footnote{C-states that further reduce idle power at the \hjj{\emph{package-level}} (e.g., C8) 
take longer to transition and require \jk{longer 
residency
times}
~\cite{gough2015cpu,haj2018power,antoniou2022agilepkgc}.} \hy{and our new proposed idle states:} \hg{$C6A$ and $C6AE$ (\hpp{which replace $C1$ and $C1E$; see Sec.~\mbox{\ref{sec:technique}}})}. The Table shows for each state and two frequency levels (base: $P1$, and minimum: $Pn$) the power \jk{consumption} of \jk{a} modern server CPU core.\footnote{\hg{The \jk{microsecond-scale}
transition times in Table \mbox{\ref{tab:c-states}} \hpp{represent the worst-case software+hardware entry+exit latency (to start executing the first instruction) and not the actual hardware transition latency}}\mbox{\cite{CPU_idle}}\hg{. For example, the hardware \jk{transition latency for the} C1 C-state is only a few nanoseconds (cycles) since C1 mainly performs clock-gating (Fig. \mbox{\ref{fig:c1_c6_c_state_flow}}(a))}\mbox{\cite{rogers2012core,schone2015wake,schone2019energy,intel_atom_2010}}.\label{fn_tran_lat}}
Clearly, the C-state that a CPU core resides \hjj{in} 
determines \jk{the core's} power.
Transitioning to a deeper (or shallower) C-state 
incurs a transition latency during which the
core cannot perform useful work.
Consequently, power management controllers
only switch to a deeper C-state if they predict that waking-up will not be needed before a target residency time.

\begin{table}[h]
\small
\centering
\caption{C-states available on the Intel Skylake server (SKX) core \cite{intel_idle} \hy{and  \mbox{\tech's} new $C6A$ and $C6AE$ C-States.}}
\label{tab:c-states}
\resizebox{\linewidth}{!}{%
\begin{tabular}{llll}
\hj{Core} C-state & Transition time\textsuperscript{\ref{fn_tran_lat}} & \hj{Target} residency time & Power per core \\
\cmidrule(lr){1-1} \cmidrule(lr){2-2} \cmidrule(lr){3-3} \cmidrule(lr){4-4}
$C0$ ($P1$)    & N/A        & N/A         & ${\sim}4$W \\
$C0$ ($Pn$)    & N/A        & N/A         & ${\sim}1$W \\
$C1$ ($P1$)   & $2${\textmu}s  & $2${\textmu}s   & $1.44$W \\
\hy{$C6A$ ($P1$)}    & \hy{$2${\textmu}s} & \hy{$2${\textmu}s} & \hy{${\sim}0.3$W}\\
$C1E$ ($Pn$)   & $10${\textmu}s & $20${\textmu}s & $0.88$W \\
\hy{$C6AE$ ($Pn$)}    & \hy{$10${\textmu}s} & \hy{$20${\textmu}s} & \hy{${\sim}0.23$W}\\
$C6$    & $133${\textmu}s & $600${\textmu}s & ${\sim}0.1$W 
\end{tabular}%
}
\end{table}

Under \jk{commonly-used} C-state transition policies, servers running latency-critical services rarely enter a deep idle power state (e.g., C6) because:
1) residency time is hard to guess, as the duration of busy/\jk{idle} periods is irregular; and
2) stringent service latency requirements ($30$--$250${\textmu}s~\cite{chou2019mudpm,zhan2016carb}) cannot be met when transitioning out of \jk{a deep C-state} requires \hjj{tens or hundreds of} microseconds.
As a result, idle CPU cores only briefly enter shallow C-states \jk{(e.g., C1)}, 
with limited power savings.



We claim that the inefficiency of the C-state hierarchy \jk{with respect to microsecond-latency} events is not fundamental, but a byproduct 
of being \emph{oriented for client} systems.
Major server vendors often design a base microarchitecture upon which both client and server CPUs are built. For example, Intel CPU core design is a single development project where client and server processors are based on the same master design~\cite{Skylake_die_server,kumar2017intel}.
Within such \jk{a} project,
energy 
optimizations are mostly \jk{targeted towards} client CPUs, which are used in battery-operated devices, while server CPUs are mostly optimized for performance.
Therefore, features such as C-states are designed for client applications (e.g., video playback, conferencing, gaming~\cite{19_MSFT,mobilemark}), which, \jk{in contrast} to latency-critical microservices, typically present long and predictable idle periods, allowing processors to exploit \jk{existing} deep package C-states (i.e., $C7$, $C8$, $C9$, and $C10$) \cite{kurd2014haswell,chi2015low,haj2020sysscale,haj2020techniques} \jk{with even larger transition latencies}.
For example, a client CPU spends ${>}80\%$ of video streaming time in the $C8$ package C-state~\cite{haj2021burstlink}.

Prior work (\autoref{sec:related}) proposes 
management techniques to enable
datacenter processors to leverage \jk{existing} deep C-states effectively \hjj{(without changing the C-state architecture)}.
In contrast, our \textbf{goal} is to directly address the root cause of the inefficiency, namely the high transition latency (tens \hjjj{or hundreds} of microseconds; see~\autoref{tab:c-states}) to/from deep C-states.
We propose \emph{\techL(\tech)}\jk{,} 
a new deep C-state architecture optimized for 
processors in modern datacenters running user-facing \jk{latency-sensitive} workloads.
\tech markedly reduces the transition latency of deep idle power states, while retaining most of their power savings, making deep C-states usable \jk{in such datacenter services}. \tech redesigns the state-of-the-art CPU core deep C-states based on
\textbf{three 
power management techniques}.
First, instead of 
shutting off the core power 
when entering a deep C-state, \tech uses medium-grained power-gates distributed across the core and maintains the core context in the power-ungated domain.
This approach shaves off several microseconds 
by removing the need to save and restore the context.
Second, instead of shutting down private caches (i.e., L1 and L2), which requires flushing and adds several microseconds to the transition latency, \tech keeps them power-ungated, \jk{along with a small} control logic for cache coherence.
\tech implements a sleep mode in caches 
that helps reduce the core voltage to a minimum operational level and limit leakage power 
of the power-ungated domain.
Third, instead of shutting down the clock distribution, 
\tech clock-gates the core components and clock distribution, while keeping the power-efficient all-digital phase-locked loop (ADPLL \cite{fayneh20164}) clock \ho{generator on} %
   and locked. 
Keeping the APLL on
shaves few more microseconds of transition latency at a minimal extra idle power consumption.

While we demonstrate \jk{the potential of AgileWatts} for Intel server processors, which represent more than $80\%$ of the server processor market~\cite{intel_amd_marketshare},
our proposed techniques are general and applicable to most server processor architectures.

\hp{This work makes the following \textbf{contributions}:}
\begin{itemize}[leftmargin=*,noitemsep,topsep=0pt]
\item To our knowledge, \techL(\tech) is the first practical highly-efficient core C-state architecture, directly targeting the \jk{energy inefficiency of} killer microseconds for datacenter servers running latency-critical applications.
\item 
\tech dramatically reduces the transition latency of deep idle states while \jk{keeping almost all} of their power savings.
\item \tech architecture employs medium-grained power gating and voltage control to reduce the need for saving/restoring the microarchitectural context and flushing \jk{caches. As a result, \tech saves} several tens of microseconds of transition latency \jk{to/from a deep idle C-state}. 
\item \jk{Our evaluation} 
shows that \tech  \jk{significantly reduces} (\hjj{by} up to $71\%$) the energy consumption of the evaluated services. \tech's new deep C-states have up to $900\times$ faster transition latency than \jk{that of} the \jk{existing} deepest idle core c-state $C6$ while their power \jk{consumption} is only $5\%$-$7\%$ of \jk{that of} the active state (C0). 

\end{itemize}

%% file: _02_motivation.tex
\section{Motivation}
\label{sec:motivation}

\newcommand\CZeroW[0]{$4$W\xspace}
\newcommand\COneW[0]{$1.44$W\xspace}
\newcommand\COneT[0]{$2\mu$s\xspace}
\newcommand\CSixW[0]{$0.1$W\xspace}

Before diving into details, we analyze the opportunity of a new, agile deep idle state for datacenter processors.

It is well known that servers running latency-critical applications usually operate at low utilization $5\%$--$25\%$~\hjj{\cite{lo2014towards,B1,B2,B3,B4,meisner2009powernap}} to \hj{keep} tail latency \hj{under control}. It is also understood that \hj{dynamic} workload  behavior prevents modern cores from entering deep idle states during \hj{idle} periods \hj{of such application}. Previous work \hj{shows} that, for a key-value store (e.g., Memcached \cite{jose2011memcached}) workload, cores never enter a C-state deeper than $C1$ (the shallowest C-state, see \autoref{tab:c-states}) when running at $20\%$ or higher load~\cite{chou2016dynsleep, lo2014towards} (our experiments in \autoref{sec:evaluation} confirm this).
A search workload is
slightly more efficient, with cores reaching deeper C-states $20\%$ of the time at $25\%$ load, but only $5\%$ of the time at $50\%$ load, thus still spending $55\%$ and $45\%$ of the time in $C1$, respectively~\cite{lo2014towards}.
These examples point to a large opportunity for a new C-state that has \hj{a transition} latency similar to $C1$ but much lower idle power than the \COneW of $C$1.

Since the deepest C-state is $C6$ (\autoref{tab:c-states}), we 
estimate an upper bound of the average power (AvgP) savings
for the ideal case of a deep idle state with the 
latency of $C1$ (i.e., \COneT) and the 
power 
of $C6$, i.e., \CSixW per core, using Eq.~\ref{eq:ideal-savings}.
%
\begin{align}
AvgP_{baseline} &= \textstyle\sum_{i \in \{0, 1, 6\}}\left( R_{C_i} \times P_{C_i} \right) \nonumber\\
AvgP_{savings} &= R_{C_1} \times (P_{C_1} - P_{C_6}) \nonumber\\
AvgP_{savings\%} &= (AvgP_{savings} / AvgP_{baseline}) \times 100 \label{eq:ideal-savings}
\end{align}
%
$R_{C_i}$ denotes the residency at power state $Ci$, i.e., the fraction of 
time a CPU core spends in state $Ci$. $P_{C_i}$ denotes the average CPU core power 
in state $Ci$ (reported in~\autoref{tab:c-states}).

Referring to our examples from prior work \hj{\cite{jose2011memcached,chou2016dynsleep,lo2014towards}}, 
given 1) the C-state residencies for the search workload at $50\%$ and $25\%$ \hj{loads} (i.e., $R_{C_0}$\,=\,$50\%$, $R_{C_1}$\,=\,$45\%$, $R_{C_6}$\,=\,$5\%$ and $R_{C_0}$\,=\,$25\%$, $R_{C_1}$\,=\,$55\%$, $R_{C_6}$\,=\,$20\%$), and for the key-value store at $20\%$ load (i.e., $R_{C_0}$\,=\,$20\%$, $R_{C_1}$\,=\,$80\%$, $R_{C_6}$\,=\,$0\%$), and 2) C-states power 
from \autoref{tab:c-states}:
then there is potential for a
$23\%$, $41\%$, and $55\%$ reduction in core power for the three \hj{loads}, respectively. Lighter loads can have even \emph{higher} power savings.


The rest of the paper illustrates how \tech~\hjj{can enable} a large part of this \hj{substantial} power-saving opportunity by defining a new low-latency deep idle state we call \hjj{\emph{\CAgile} (\emph{C6 Agile})}.

%% file: _03_background.tex
\section{Background}
\label{sec:bg}

We provide a brief overview of different power management components and techniques used in modern processors.

\begin{figure}[ht]
    \centering
	\includegraphics[trim=0.5cm 0.5cm 0.5cm 0.5cm, clip=true,width=0.85\linewidth,keepaspectratio]{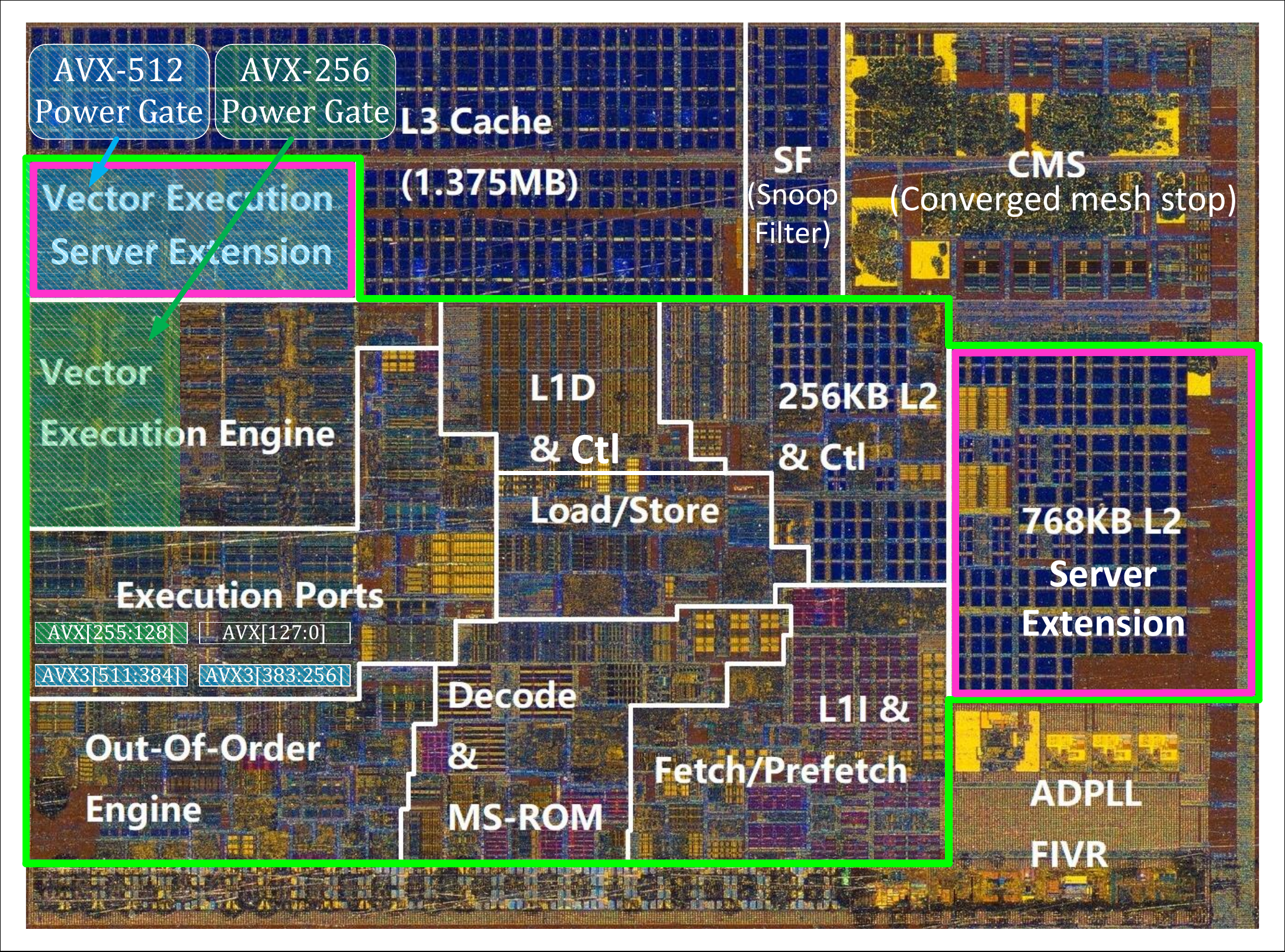}
	\caption{A\hj{n Intel} Skylake server
	core slice 
	~\cite{skx_die_annotations}.
	The core is bordered with green, and the AVX-512 and L2 extensions~\cite{kumar2017intel} (unavailable in client CPU cores) are bordered in pink. 
	The 256-bit and 512-bit AVX units have separate power gates~\hp{\mbox{\cite{mandelblat2015technology,intel_avx512,fayneh20164,haj2021ichannels}}}, as \hj{shown} in the figure.}
	\label{fig:skx_floorplan}
\end{figure}

\noindent\textbf{Server and Client Cores.}
Major server vendors have nearly the same 
core microarchitecture for client and server CPUs.
For example, Intel CPU core design is a single development project
that has two derivatives, one for server and one for client CPUs \cite{Skylake_die_server}. Fig. \ref{fig:skx_floorplan} shows the AVX and L2 extension of the Intel Skylake server core over the client core \cite{kumar2017intel}.  

\noindent\textbf{Clock Distribution Network (CDN).}
A CDN distributes the signals from a common point (e.g., clock generator) to all the elements in the system that need it. 
Modern processors use an all-digital phase-locked loop (ADPLL) to generate the CPU core clock~\cite{tam2018skylake}. An ADPLL maintains high performance with significantly less power as compared to conventional PLLs. For example, the power of an ADPLL in Skylake, shown \hj{at the} bottom \hj{right} of Fig.~\ref{fig:skx_floorplan}, is only $7mW$ at $4GHz$~\cite{fayneh20164}.



\noindent\textbf{Power Delivery Network (PDN).}
The three commonly-used PDNs in modern CPUs are: 1) \lowercase{\emph{Integrated Voltage Regulator}} (IVR)~\cite{burton2014fivr,nalamalpu2015broadwell,tam2018skylake,icelake2020},
2) \lowercase{\emph{Motherboard Voltage Regulator}} (MBVR)~\cite{rotem2011power,fayneh20164,haj2019comprehensive}, 
and 3) \lowercase{\emph{Low Dropout Voltage Regulator}} (LDO VR)~\cite{singh20173,singh2018zen,burd2019zeppelin,beck2018zeppelin,toprak20145}.
For example, recent Intel server CPUs implement a fully\hj{-}integrated voltage regulator (FIVR) per core, as shown at the bottom \hj{right} of Fig.~\ref{fig:skx_floorplan} ~\cite{fayneh20164,Skylake_die_server}.

\begin{figure}[th]
    \centering
	\includegraphics[trim=0.5cm 0.5cm 0.5cm 0.5cm, clip=true,width=0.9\linewidth,keepaspectratio]{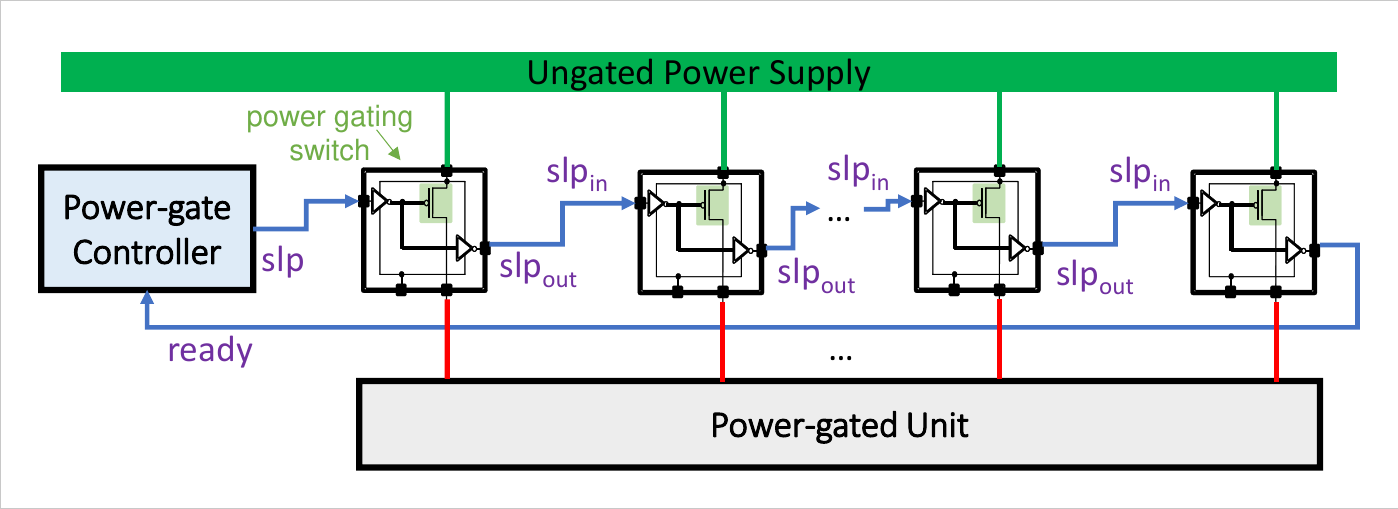}
	\caption{Staggered power-gate wake-up by daisy-chaining the control signals of the power-gating switches.}
	\label{fig:staggered_pg}
\end{figure}

\noindent\textbf{Staggered Power-gate Wake-up.}
Power-gating is a 
technique that is used to eliminate leakage of idle circuits~\cite{hu2004microarchitectural,haj2018power,gough2015cpu,yahya2022darkgates}. Typically, the wake-up latency from a {power-gated} state requires a few to tens of cycles~\cite{kahng2013many,gough2015cpu}. However, to reduce the worst-case peak in-rush current~\cite{chadha2013architectural,usami2009design,agarwal2006power,abba2014improved} and voltage-noise in the PDN (e.g., di/dt noise~{\cite{larsson1997di,gough2015cpu,haj2018power}}) when waking up a power-gate, {a {power-gate} controller uses} a \emph{staggered} wake-up technique~\cite{chadha2013architectural,usami2009design,agarwal2006power}, shown in Fig.~\ref{fig:staggered_pg}.

The technique turns on different power-gate switch cells in \hj{a staggered manner}\hjj{,} to limit the current spike from the power supply. To do so, the \hj{input sleep} \hj{(}\texttt{$slp_{in}$}\hj{)} and \hj{output sleep} \hj{(}\texttt{$slp_{out}$}\hj{)} \hj{signals} of the switch cells are daisy-chained. The controller issues a signal to the first \texttt{$slp_{in}$}, and it receives an acknowledgement (\texttt{$ready$}) from the last \texttt{$slp_{out}$}, indicating that the power-gate is fully conducting. An alternative wake-up technique groups switch cells into multiple chains each in a daisy-chain configuration. 
Doing so allows the power-gate controller to tune (e.g., post-silicon) a unit's wake-up time by controlling the assertion time for each chain.     
Modern CPUs implement the staggering  technique \hp{\mbox{\cite{kahng2013many,akl2009effective,kahng2012tap}}}; e.g., the Intel Skylake core staggers the wake up of the AVX power-gates over $15ns$ to reduce in-rush current \hp{\mbox{\cite{fayneh20164}\cite[Sec. 5]{haj2021ichannels}}}.

\noindent\textbf{Core C-states.}
Power saving states enable cores to reduce their \hj{power} consumption during idle periods.
Modern processors support various C-states,
for example, Intel's Skylake architecture offers the following four: $C0$, $C1$, $C1E$, $C6$ ~\cite{gough2015cpu,haj2018power,haj2018energy}. Table~\ref{tab:c_state_actions} describes the state for various \hj{core components} for each existing core C-state as well as \hjj{in} our proposed idle states: \hg{$C6A$ and $C6AE$ (which replace $C1$ and $C1E$, Sec.~\mbox{\ref{sec:technique}})}.
While C-states reduce power \hj{consumption}, during the entry-to and exit-from a C-state 
a core cannot be \hj{used}. For example, it is estimated that $C6$ requires $133${\textmu}s of transition time (Table \ref{tab:c-states}). As a result, entry-exit latencies can degrade the performance of services that have microsecond\hj{-level} processing latency \hjj{requirements}, such as in user-facing applications~\cite{jose2011memcached}.

\begin{table}[ht]
\centering
\caption{\hj{Skylake server core components' states in} core C-states\hj{,}  \hy{including \mbox{\tech's} new $C6A$ and $C6AE$ C-States \cite{oneintel}.}}
\label{tab:c_state_actions}
\resizebox{\linewidth}{!}{%
\begin{tabular}{llllll}
\textbf{C-State} & \textbf{Clocks} & \textbf{ADPLL} & \textbf{L1/L2 Cache}   & \textbf{Voltage}          & \textbf{Context}      \\ 
\cmidrule(lr){1-1} \cmidrule(lr){2-2} \cmidrule(lr){3-3} \cmidrule(lr){4-4} \cmidrule(lr){5-5} \cmidrule(lr){6-6}
\textbf{C0}         & Running     & On  & Coherent     & Active           & Maintained   \\
\textbf{C1}         & Stopped     & On  & Coherent     & Active           & Maintained   \\
\hy{\textbf{C6A}}    & \hy{Stopped}& \hy{On} & \hy{Coherent} & \hy{PG/Ret/Active}     & \hy{In-place S/R} \\
\textbf{C1E}         & Stopped     & On  & Coherent     & Min V/F       & Maintained   \\
\hy{\textbf{C6AE}}    & \hy{Stopped}& \hy{On} & \hy{Coherent} & \hy{PG/Ret/Min V/F}     & \hy{In-place S/R} \\
\textbf{C6}         & Stopped     & Off & Flushed & Shut-off        & S/R SRAM 
\end{tabular}%
}
\end{table}

\begin{figure}[ht]
    \centering
	\includegraphics[trim=0.5cm 0.5cm 0.5cm 0.5cm, clip=true,width=0.9\linewidth,keepaspectratio]{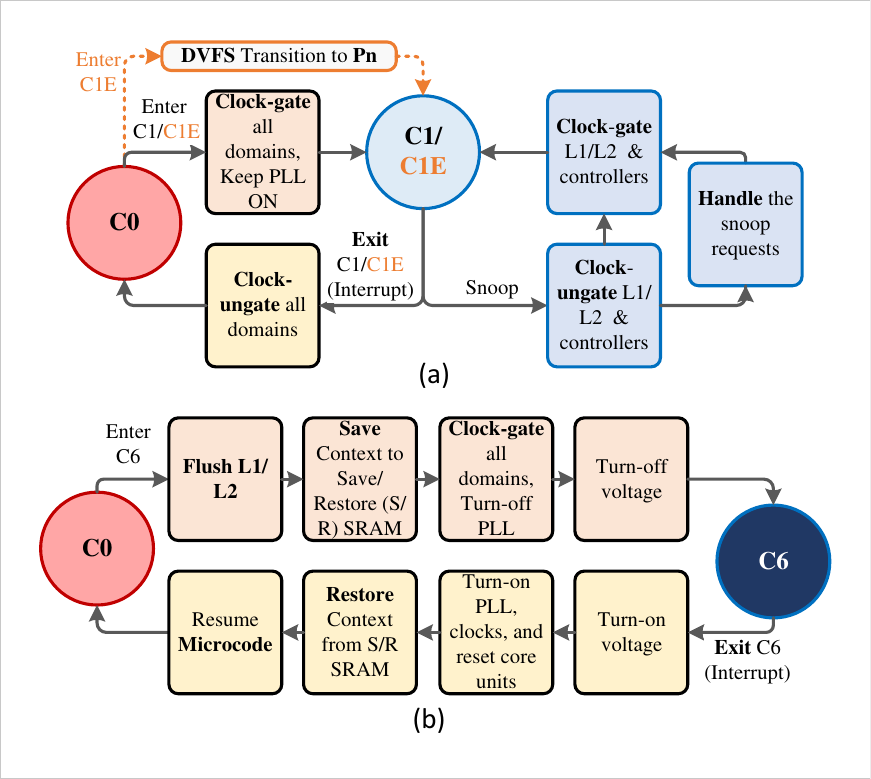}
	\caption{Entry and Exit Flows for (a) C1 and (b) C6.}
	\label{fig:c1_c6_c_state_flow}
\end{figure}

\noindent\textbf{Core C-state Entry and Exit Flow\hj{s}.} The C1/C1E, and C6 entry and exit flows are shown in Figs. \ref{fig:c1_c6_c_state_flow}(a) and \ref{fig:c1_c6_c_state_flow}(b), respectively, and are discussed in detail in \cite{rogers2012core,schone2015wake,schone2019energy,intel_atom_2010}.

\noindent\textbf{Core C6 Entry/Exit Latency.}
We analyze the $C6$ entry/exit latency based on \hj{an} x86  implementation~\cite{rogers2012core}. 
$C6$ entry latency is \emph{dominated by the L1/L2 cache flush} \hj{time}. This flush time varies depending on 1) the \hjj{fraction of} \hj{cache}
lines \hjj{that are dirty} and 2) the core frequency\hjj{,} when entering C6; e.g., flushing a $50\%$ dirty cache at 800MHz takes ${\sim}75${\textmu}s. 
The time to transfer core state to/from the save/restore (\emph{S/R}) SRAM depends on the core clock frequency; e.g., at 
$800MHz$, the latency is ${\sim}9${\textmu}s. Including control flow overhead and the power-gate controller \hj{latency}, the overall CPU core $C6$ entry time is ${\sim}87${\textmu}s.

$C6$ exit latency is significantly faster, taking ${\sim}30${\textmu}s from the wake-up interrupt to resuming \hj{core} execution. This latency includes ${\sim}10${\textmu}s for hardware wake-up, including power-ungating, PLL relock, reset, and fuse propagation. State and microcode restoration take\hj{s} ${\sim}20${\textmu}s~\cite{rogers2012core,schone2015wake,schone2019energy,intel_atom_2010}.




%% file: _04_technique.tex
\section{\techL ({\tech}) Architecture}
\label{sec:technique}

\newcommand\ufpg[0]{\emph{Units' Fast Power-Gating}\xspace}
\newcommand\UFPG[0]{UFPG\xspace}
\newcommand\ccsm[0]{\emph{Cache Coherence and Sleep Mode }\xspace}
\newcommand\CCSM[0]{CCSM\xspace}


\tech introduces a new \emph{core deep idle power state}, C6A (C6 Agile), with \emph{close to zero\hj{-}Watt} power \hj{consumption} and \emph{nanosecond-scale} entry/exit latency.
Thanks to its low latency, servers running latency-critical applications can enter $C6A$ during short and irregular idle periods, unlocking significant energy savings.
Additionally, \tech defines C6AE (C6A Enhanced, analogous to $C1E$), a lower-power variant of C6A
that further reduces leakage power
by lowering core voltage to a minimum operational level.
Our discussion focuses on the C6A design and operation and points out C6AE differences when relevant.

The C6A state is based on two key ideas: \ufpg~(\UFPG) (discussed in \autoref{sec:UFPG}) and a \ccsm~(\CCSM) (discussed in \autoref{sec:CCSM}).
The new power management \hj{flow} that coordinates the \UFPG and the \CCSM at nanosecond granularity is presented in \autoref{sec:c6a_low}.



\subsection{Units' Fast Power-Gating~(\UFPG)}
\label{sec:UFPG}
\tech UFPG is a low-latency power-gating (PG) architecture that shuts off most of the core units while retaining the context in place, thus, enabling a transition latency of tens of nanoseconds.
Conventional context retention techniques (see C6 C-state flow in \autoref{fig:c1_c6_c_state_flow}(b)) sequentially \hj{saves}/\hj{restores} the context to/from external SRAM before/after power-gating/un-gating~\cite{rogers2012core,haj2021ichannels,fayneh20164,haj2020techniques}. This process adds several \hj{(e.g., $5$--$10${\textmu}s)} microseconds
to the entry/exit latency.
Instead, \tech retains the context in place, completely removing that overhead at a very small additional idle power cost.

\tech enables in-place context retention with a \emph{medium-grain} PG approach.
This is in contrast to the \emph{coarse-grain} PG used in Skylake client cores, where the entire core is under the same power gate~\hp{\cite{rotem2011power,10_jahagirdar2012power,fayneh20164,12_howse2015tick}} and context is saved/restored externally.
Our approach leverages the same ideas used by the PG for the AVX-256 and AVX-512 core units~\hp{\mbox{\cite{mandelblat2015technology,intel_avx512,fayneh20164,haj2021ichannels}}} in recent server and client cores (see \autoref{fig:skx_floorplan}).
These PG techniques  require \hj{only} $10$ to $20$ nanoseconds to power-gate/un-gate a unit~\hp{\cite{fayneh20164} \cite[Sec. 5]{haj2021ichannels}} because they power and retain the unit's context in place and avoid having to save and restore it externally.

\begin{figure}[ht]
    \centering
	\includegraphics[trim=0.5cm 0.6cm 0.5cm 0.6cm, clip=true,width=0.85\linewidth,keepaspectratio]{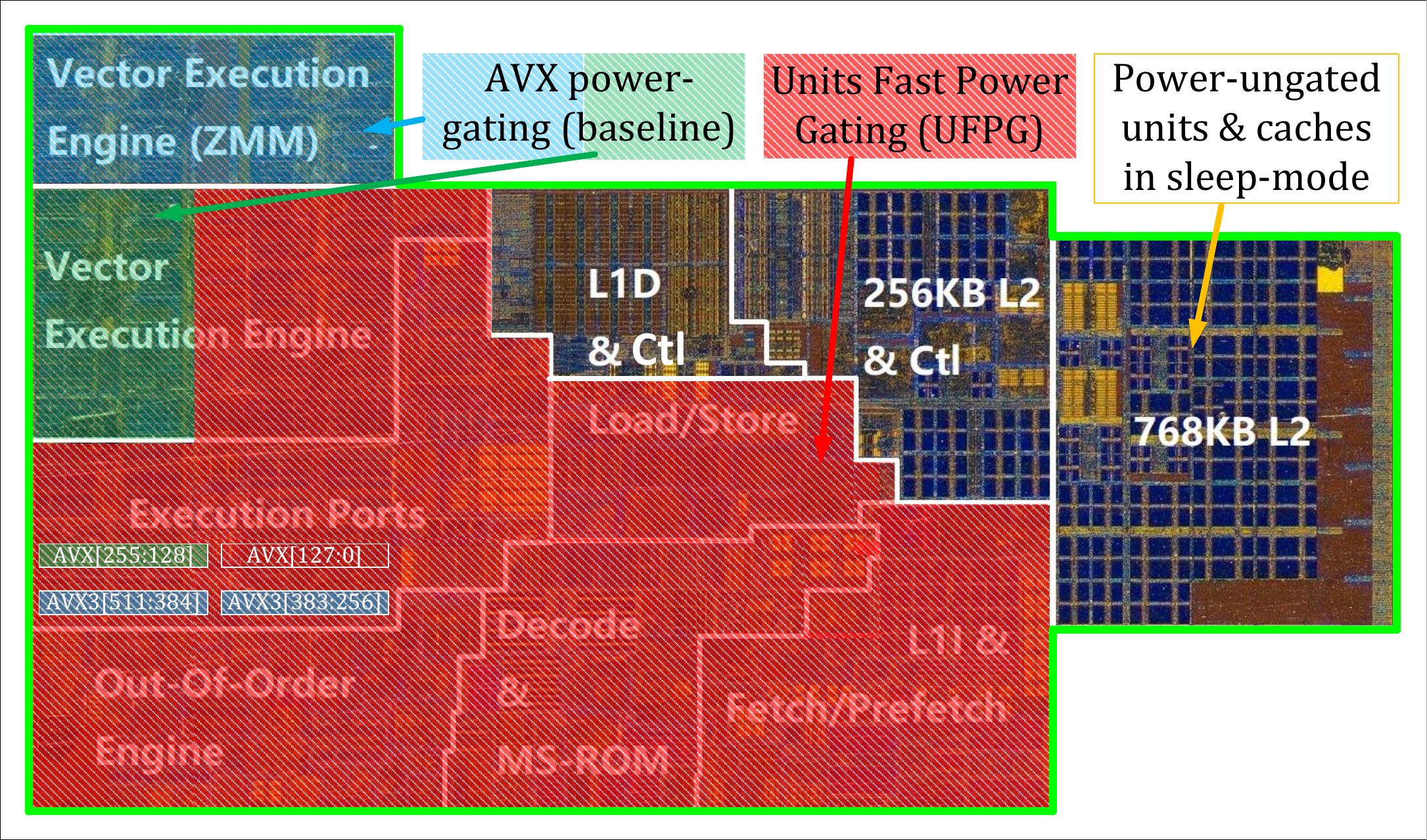}
	\caption{Medium-grain PG for the majority (area shaded in red) of the core units, excluding the L1 and L2 caches and their controllers.}
	\label{fig:skx_ufpg}
\end{figure}

\tech's \emph{medium-grain} power gating applies to the majority of the core units (shaded red in \autoref{fig:skx_ufpg}) and excludes the L1 and L2 and their controllers (handled separately in \autoref{sec:CCSM}).

Within the medium-grain PG region, \tech leverages multiple techniques to retain the context in place, and enable fast (\hj{several-}nanosecond) transition latency.
The context of a modern CPU core 
is ${\sim}8$kB\footnote{\hy{\mbox{\cite{haj2020techniques}} \hj{shows} that the context saved/restored for an entire Skylake client mobile SoC is ${\sim}200$KB. This includes the context of \hj{four cores}, \hj{an} integrated GPU, \hj{the} uncore, and \hj{the} system agent. We \hj{estimate} the single-core context 
based on a core's relative die area, showing  ${\sim}8$kB context, similar to previous Intel references \cite{gerosa2008sub,jahagirdar2012method}. 
}} 
(estimated as the amount of state that C6 saves)~\cite{gerosa2008sub,jahagirdar2012method} and falls into two categories:
i) \textit{registers}, such as configuration and status registers (CSRs) or fuse registers, and
ii) \textit{SRAMs}, such as firmware persistent data and patches~\cite{haj2020techniques,rogers2012core}.
We discuss next three 
techniques \tech uses to 
efficiently retain the context during C6A; the first two apply to registers and the third to SRAM.

\subsubsection{Placing Unit Context in the Ungated Domain}
\label{sed:ungated-domain}
\begin{figure}[ht]
    \centering
	\includegraphics[trim=0.85cm 0.8cm 0.6cm 0.6cm, clip=true,width=0.99\linewidth,keepaspectratio]{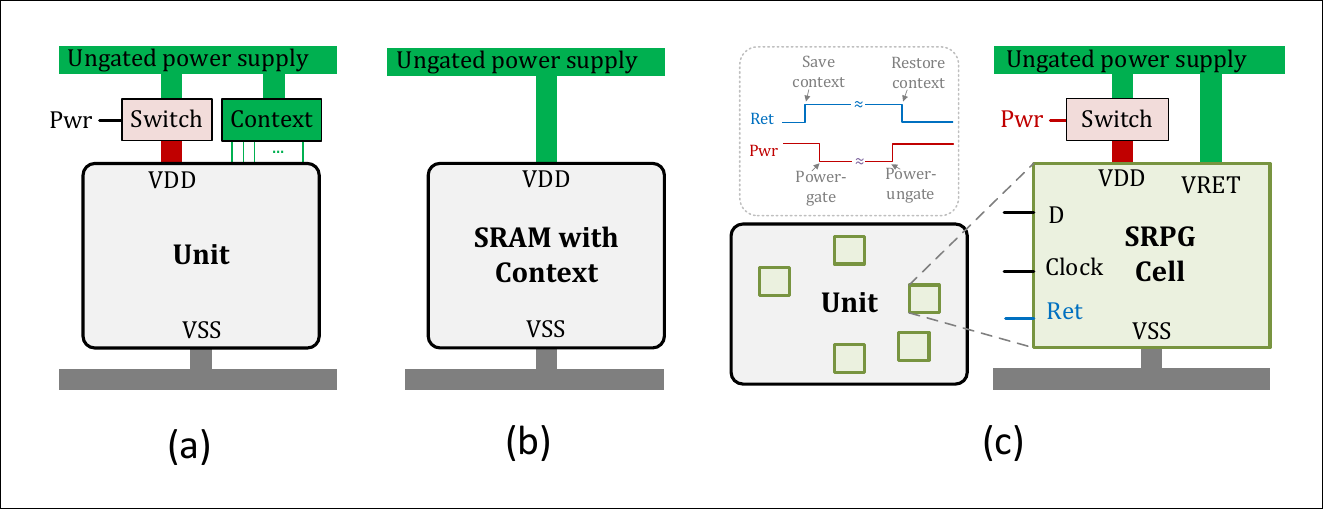}
	\caption{Context retention techniques \tech uses when \hj{power-gating} a unit: (a) Placing context in the \hb{core ungated power domain}; (b) placing SRAM with context (e.g., microcode patch) in the \hb{core ungated power-domain}; (c) using 
	SRPG \hj{cells} for distributed context.}
	\label{fig:state_retention}
\end{figure}

One option to retain the context of a power-gated unit is to place its registers outside the power-gated region, i.e., in the \hb{core's ungated domain}, as shown in \autoref{fig:state_retention}(a). This is suitable for units with small context
(e.g., execution units); 
Intel likely uses this technique for the AVX execution units~\cite{mandelblat2015technology,lackey2002managing}.
\tech \ho{uses this 
technique for all core units that require only a local context to be retained, i.e., this is not applicable to a unit with a distributed context that is impractical to relocate to a centralized un-gated region}.
The following units satisfy this requirement: 1) all execution units (besides AVX), 2) execution ports, and 3) the out-of-order engine.

\subsubsection{State Retention Power Gates (SRPGs)}
\label{sed:srpgs}
Moving distributed or large context to a separate un-gated area is impractical (e.g., due to timing and wiring constraints).
For this reason, \tech employs a different retention technique -- SRPGs -- for units that contain such context.
As \autoref{fig:state_retention}(c) illustrates, SRPG (i.e., \hjj{a} retention \hjjj{flip-flop}) \hjj{is a} special \hj{flip-}flop fed with two supplies: power-gated and \hb{power-ungated}.
\hjj{Such a \hjjj{flip-}flop} typically \hjj{contains} a shadow \hjj{flip-}flop to retain \hjj{its} state 
when the unit \hjj{it resides} in is power-gated~\cite{mahmoodi2004data,rabinowicz2021new,lackey2002managing}.
For example, Intel uses this technique in the chipset to retain the state of autonomously power-gated units~\cite{mandelblat2015technology}.

\subsubsection{Place SRAM Context in Ungated Power Supply}
\label{sed:ungated-sram}
Part of the CPU core context is located in SRAMs~\cite{haj2020techniques}.
While the microcode firmware is stored in read-only memory, known as microcode sequencer ROM (MS-ROM), microcode patches and data are stored in a ${\sim}2KB$ SRAM~\cite{gwennap1997p6,ermolov2021undocumented}.
This SRAM is initialized at boot time and should be retained when power-gating the microcode unit.
The C6 exit flow 
re-initializes the content of this SRAM from core's S/R SRAM in a separate un-gated uncore domain; this 
process is sequential and can take \hj{several} microseconds \cite{haj2020techniques,rogers2012core}.
\tech avoids the need to re-initialize the microcode patch SRAM by powering it with a separate \hb{core un-gated supply}\hj{,} as shown in \autoref{fig:state_retention}(b).

\subsection{Cache Coherence and Sleep Mode~(\CCSM)}
\label{sec:CCSM}
To avoid the high latency (tens of microseconds) to flush private caches (i.e., L1D and L2) \hj{in order} to power-gate them, \tech instead keeps them power-ungated (see~\autoref{fig:skx_ufpg}) when transitioning to C6A.
This has two design implications: first, \tech needs to employ other power-saving techniques to reduce the power of the cache domain; and second, a core in C6A state still needs to serve \hj{coherence} requests (i.e., snoops)
~\cite{molka2009memory,hackenberg2009comparing}.

\tech employs two key techniques to reduce the power \hj{consumption} of the power-ungated private cache domain.
First, unless a \hj{coherence} request is being served, \tech keeps this domain clock-gated to save its dynamic power.
Second, \tech leverages the \emph{cache sleep-mode} technique~\cite{huang2013energy,flautner2002drowsy,chen201322nm,rusu20145,rusu201422}, which adds sleep transistors to the SRAM \hj{arrays} of private caches.
These sleep transistors reduce the SRAM array's supply voltage to the lowest level that can safely retain the SRAM content while significantly reducing leakage power.

Since private caches are not flushed when a core enters C6A, \tech must allow the core \ho{to} respond to snoop requests
~\cite{gough2015cpu,haj2018power,rotem2011power}.
\tech keeps the logic required to handle cache snoops in the power-ungated (but clock-gated) domain together with the private caches.
It also \hj{uses} minimal logic (\emph{same logic used in C1}) to detect incoming snoop requests in an always-active (i.e., neither power-\hjj{gated} \hj{nor} clock-gated) domain.
As soon as this 
logic detects incoming snoop traffic, it temporarily increases the SRAM array voltage through the sleep transistors and reactivates the clock of the private caches for the time required to respond to snoop requests.

\subsection{C6A Power Management Flow}
\label{sec:c6a_low}
\begin{figure}[ht]
    \centering
	\includegraphics[trim=0.65cm 0.6cm 0.6cm 0.6cm, clip=true,width=0.99\linewidth,keepaspectratio]{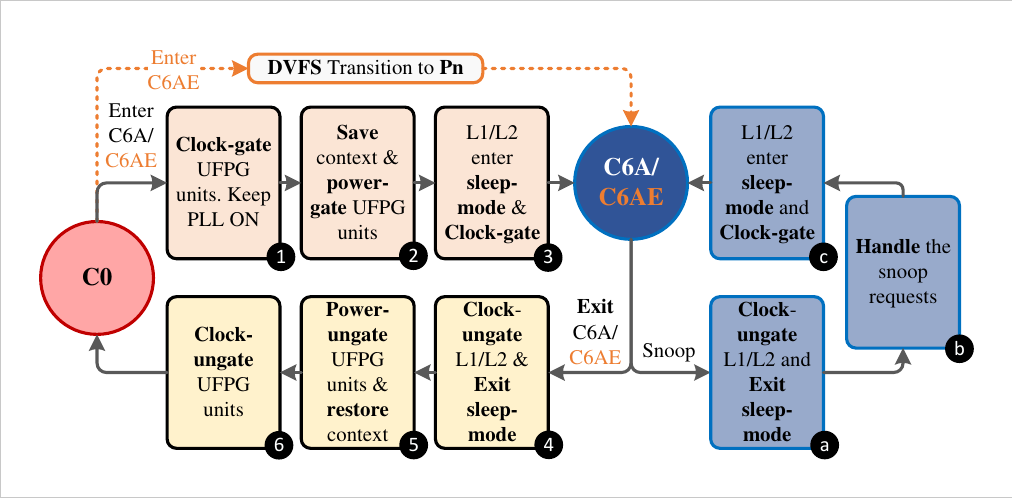}
	\caption{Power management flows for the C6A\hj{/C6AE states.}}
	\label{fig:c6a_c_state_flows}
\end{figure}

\tech implements the C6A 
flow within the core power management agent (PMA)~\cite{rotem2012power}.
This flow, shown in \autoref{fig:c6a_c_state_flows}, orchestrates the transitioning between the C0 and C6A C-states and handles \hj{coherence} traffic while in \hj{the} C6A state.

Similar to other C-states, the operating system triggers C6A \emph{entry} by executing the MWAIT instruction \cite{gough2015cpu,haj2018power}.
The first step \circledb{1} in the entry flow clock-gates the \UFPG domain (\autoref{sec:UFPG}), while keeping the core phase-locked loop (PLL) powered-on.
When entering C6AE, the PMA additionally initiates a non-blocking transition to Pn -- the P-state with lowest frequency and voltage.
Subsequently \circledb{2}, the flow saves (in place) the \UFPG domain context and shuts down its power.
Finally \circledb{3}, the flow sets the private caches into \emph{sleep mode} (\autoref{sec:CCSM}) and shuts down their clock.
After these three steps\hj{,} the core is in  $C6A$ (or $C6AE$) state.

When a \emph{snoop} request arrives while the core is in the $C6A$ (or $C6AE$), the PMA temporarily activates the private caches to respond.
First, \circleda{a} the flow clock-ungates the private cache domain and adjusts its supply voltage to exit \emph{sleep mode}.
At this point \circledb{b}, the caches can handle the snoop requests.
Finally \circleda {c}, when all outstanding snoop requests are serviced, 
PMA rolls back the changes in reverse order and brings the core back into \hj{the} full $C6A$ (or $C6AE$) state.

When an \emph{interrupt} occurs, the core \emph{exits} from $C6A$ (or $C6AE$) and goes back into $C0$ (active) state.
The exit flow is simply the reverse process of the entry flow.
First \circledb{4}, the flow clock-ungates L1/L2 and exits \emph{sleep-mode}.
Next \circledb{5}, it power-ungates the UFPG units and triggers the restore signal to the SRPG flops (\autoref{fig:state_retention}(c)).
Finally \circledb{6}, the flow clock-ungates UFPG units, bringing the core to \hj{the} $C0$ active state.

%% file: _05_implementation.tex
\section{\tech Implementation and \jk{Hardware} Cost}
\label{sec:impl}
As discussed in \autoref{sec:technique}, \tech requires 
in each CPU core: \jk{1)} the \UFPG subsystem, \jk{2)} the \CCSM subsystem, and \jk{3)} the C6A controller.
This \jk{section} discusses the implementation of each component, its \hyy{power-performance-area (PPA)}  cost, and the resulting transition latency for the new C6A and C6AE states.

\subsection{PPA Modeling Methodology}\label{sec:ppa_model_meth}
\begin{figure*}[ht]
    \centering
	\includegraphics[trim=0.65cm 0.6cm 0.6cm 0.6cm, clip=true,width=0.99\linewidth,keepaspectratio]{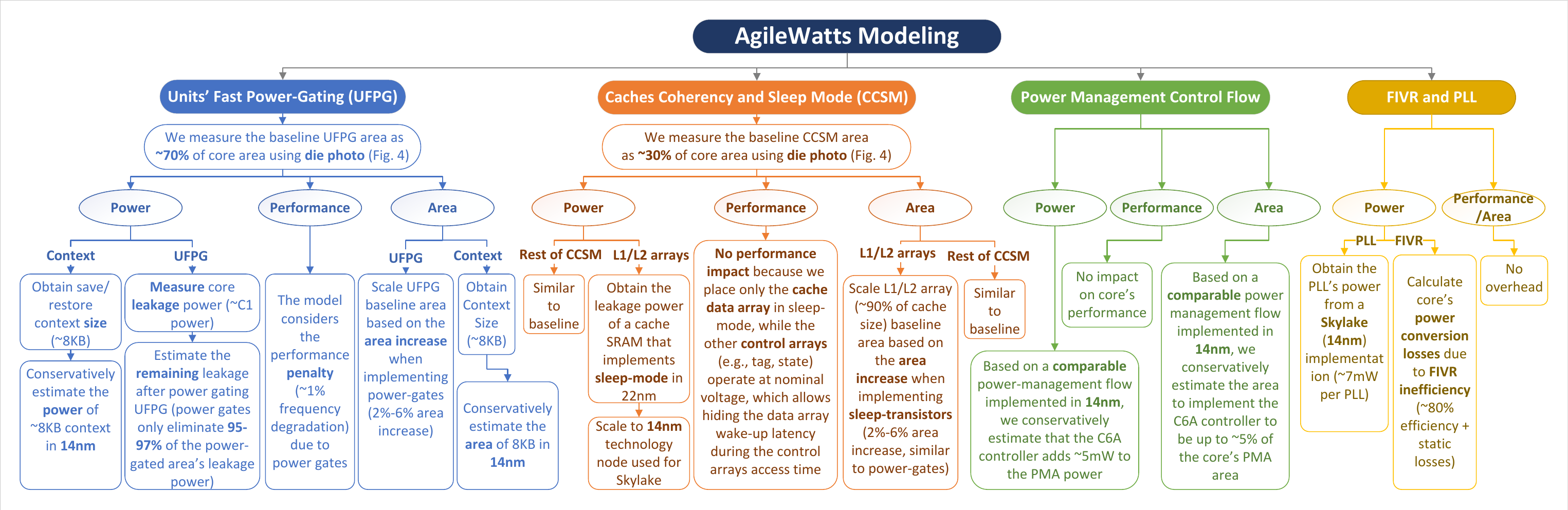}
	\caption{\hyy{\mbox{\techL} Power-Performance-Area (PPA) Modeling Methodology.}}
	\label{fig:ppa_modeling}
\end{figure*}

\hyy{\mbox{\autoref{fig:ppa_modeling}} describes at a high level \mbox{\tech's} PPA modeling methodology. In this section, we describe each of the modeling components in detail. \mbox{\autoref{tab:area_poweroverhead}} summarizes the total area overhead and power consumption of \mbox{\tech's} C6A/C6AE C-states. Our power and performance model (described in \mbox{\autoref{sec:power-model}}) uses C6A/C6AE power  and \mbox{\tech} performance overheads to estimate the average power consumption and performance impact \jk{of \tech for a given workload.}}   

\subsubsection{Units' Fast Power-Gating~(\UFPG)}\label{sec:perf_penalty}
As discussed in \autoref{sec:UFPG}, \tech's {\UFPG}  places the majority of the core units
behind power-gates that are similar to the ones used for the AVX units in recent \hj{Intel} cores.
\hy{\mbox{\tech} uses power-gates for  ${\sim}70\%$ of the core area (\hgg{measured on the die photo in} \mbox{\autoref{fig:skx_ufpg}}).

\noindent \textbf{\hyy{Power overhead.}}
The AVX power-gates and the new \UFPG, shut-off all the units in the core front-end and execution domains; however,
power-gates only eliminate $95$\,--\,$97\%$ of the leakage power~\cite{flynn2007low,petrica2013flicker,B5}, thus the \UFPG domain has residual idle power while in C6A.
Using the Intel core-power-breakdown tool~\cite{haj2016fine}, 
we derive the leakage power contribution of the power-gated units starting from the leakage power of the entire core.\footnote{\hj{The leakage of the entire core is approximately equal to the $C1$ power (\autoref{tab:c-states}), which removes \hjj{only} dynamic power by applying clock-gating.}}
Our estimation shows that the new power-gated units contribute to approximately $70\%$ of the core leakage (i.e., ${\sim}70\%$ of $C1$ power).
Hence, the 
power overhead of \UFPG (i.e., $3$--$5\%$ of the gated leakage power) is ${\sim}30$--$50mW$ at base frequency (P1), or ${\sim}18$--$30mW$ at minimum frequency (Pn).

The three \UFPG 
techniques combined retain ${\sim}8KB$  context~\cite{gwennap1997p6,ermolov2021undocumented}, which consume $0.2mW$ at retention voltage~\cite{haj2020techniques}.
To estimate the retention power at base (P1) and minimum (Pn) \jk{frequencies}, 
we conservatively multiply the retention-level power by $10\times$ and $5\times$, respectively.
Therefore, our estimate for context retention power is ${\sim}2$mW (P1) to ${\sim}1$mW (Pn).

\noindent \textbf{\hyy{Performance overhead.}}
In an active CPU core, simultaneous operations in memory or/and logic circuits demand high current flow, which creates fast transient voltage droops~\hjj{\cite{cho2016postsilicon,reddi2010voltage,thomas2016core,shevgoor2013quantifying,grochowski2002microarchitectural,gupta2008decor,haj2015compiler,leng2015gpu,reddi2009voltage,miller2012vrsync}}. 
One power-gating design challenge is the resistive voltage (IR) drop across a power gate, which exacerbates voltage droops \cite{nithin2010dynamic,radhakrishnan2021power,haj2020flexwatts,shekhar2016power,haj2019comprehensive,jotwani2010x86}.
The \jk{worst-case} voltage droop can limit the maximum attainable frequency at a given voltage since it requires additional voltage (droop) margin above the nominal voltage to enable the CPU core to run at the target frequency~\cite{nithin2010dynamic,radhakrishnan2021power,cho2016postsilicon}. An x86 implementation of CPU core power-gate leads to ${<}1\%$ frequency loss~\cite{jotwani2010x86}. \jk{Our} \tech analytical model (Sec. \ref{sec:power-model}) assumes $1\%$ \emph{frequency degradation} due to the additional CPU core power-gates (i.e., due to UFPG). 

\noindent \textbf{\hyy{Area overhead.}} 
A power gate adds $2$\,--\,$6\%$ extra \hyy{area to the gated logic (i.e., \textit{\UFPG}, \jk{covering} ${\sim}70\%$ of} core area\hyy{)}. \hyy{We conservatively use the wide overhead range (i.e., $2$\,--\,$6\%$) since}
the exact overhead depends on the specific implementation, the exact size of the gated area, the number of required isolation-cells,\footnote{Isolation-cells 
isolate the always-on units from the floating values of the power-gated units. They are typically placed on the outputs of the power-gated domains during the physical placement stage\mbox{~\cite{chadha2013architectural}.}} and technology\mbox{\cite{petrica2013flicker,ditomaso2017machine,rahman2006determination,flynn2007low,zimmer2017reprogrammable}}.}
\hy{The area overhead for the in-place \emph{context retention} techniques of the ${\sim}8$kB core context~\cite{gerosa2008sub,jahagirdar2012method} is as follows:}
\hy{First, moving the context of a unit to ungated-power typically requires ${<}1\%$ of the context area, mainly due to the isolation cells~\mbox{\cite{chadha2013architectural}}.}
\hy{Second,} the use of SRPGs for components with too large or distributed context
is a mature technique already used in products, e.g., \jk{the} Intel Skylake~\cite{mandelblat2015technology}.
Efficient SRPG designs, which use \emph{selective} context retention, require less than $1\%$ additional area relative to the power-gated area they control ~\cite{rabinowicz2021new,hyun2019allocation}.
Finally, 
\hy{including an SRAM into the ungated domain requires isolation cells that add overhead ${<}1\%$ of the SRAM area~\mbox{\cite{chadha2013architectural}}}.

\subsubsection{Cache Coherence and Sleep Mode~(\CCSM)}\label{sec:ccsm}
\tech implements  sleep-mode \jk{for private} caches similarly \jk{to} the sleep-mode used for L3 caches in multiple server products~\cite{huang2013energy,chen201322nm,rusu20145,rusu201422}. 
Cache sleep-mode implements P-type sleep transistors~\cite{huang2013energy,chen201322nm} with seven programmable settings and local bit-line float to reduce SRAM cell leakage in the data array; it also employs word-line sleep to reduce leakage further.

\noindent \textbf{\hyy{Power overhead.}} 
\emph{CCSM} implements sleep-mode using sleep-transistors in the L1/L2 SRAM data array; this technique is already used for efficient design of the L3 cache in multiple server processors in the market~\cite{huang2013energy,chen201322nm,rusu20145,rusu201422}.
We estimate the leakage of the L1/L2 SRAM data array when in sleep-mode, starting from the leakage of a $2.5$MB SRAM L3 cache with sleep-mode \hyy{that Intel implemented} at $22$nm~\cite{chen201322nm_ppt,chen201322nm}.
Based on established methodology~\cite{shahidi2018chip}, we scale\footnote{\hgg{Sec. III of \mbox{\cite{shahidi2018chip}} explains how leakage power scales with technology node \jk{size}. For a dimensional scaling factor of $\alpha$ (i.e., ${\sim}0.7\times$ when transitioning from $22$nm to $14$nm) and voltage scaling factor of $\beta$ (varies between ${\sim}0.7\times$ to a $1.0\times$; we conservatively assume no voltage scaling, i.e., $\beta=1.0$), the leakage power scales as $\alpha \beta$ (i.e., ${\sim}0.7\times$ in our case).}\label{footnote:leak_scal}} this power to the cumulative L1 and L2 capacity (${\sim}1.1$MB and $14$nm technology node used for Skylake).
The resulting leakage power estimation for the L1/L2 caches is ${\sim}55$mW.
We use the same method to estimate the power for the rest of the power-ungated units (controllers and tags), resulting in an additional $55mW$ at P1 \hpp{(i.e., in C6A)} voltage level.
Reducing the core voltage to Pn \hpp{(i.e., in C6AE)} level increases the sleep \ho{transistor %
efficiency} and reduces leakage power at Pn voltage to $40mW$. This is because a sleep transistor is effectively a linear voltage regulator (LVR). The LVR power-conversion efficiency is the ratio of the desired output voltage and the input \jk{voltage; hence,} the closer the input voltage to the output, the higher the  power-conversion efficiency~\cite{luria2016dual,huang2016fully,haj2020flexwatts}.

\noindent \textbf{\hyy{Performance overhead.}} 
In \tech, only the data array (which accounts for more than $90\%$ of the \hj{L1/L2} cache size) is placed in sleep-mode, while the other control arrays (e.g., tag, state) operate at nominal voltage.
Doing so allows \emph{hiding the data array wake-up latency} during the control \jk{array access time, thereby} eliminating any performance degradation compared to operation without the sleep mode.

\noindent \textbf{\hyy{Area overhead.}} 
\hy{Implementing sleep-mode using \jk{sleep transistors} for the SRAM data array of the private caches requires additional area similar to power-gates (i.e., $2$--$6\%$ of the SRAM area)~\mbox{\cite{petrica2013flicker,ditomaso2017machine,rahman2006determination,flynn2007low,flautner2002drowsy}}); a recent implementation reports a $2\%$ area overhead\mbox{~\cite{zimmer2017reprogrammable}}.}

\subsubsection{\tech Power Management Control Flow}
The main implementation consideration to realize the C6A flow in~\autoref{fig:c6a_c_state_flows} is a mechanism to control in-rush current~\cite{fayneh20164,jeong2012mapg}.
This needs to support staggered wake-up, so as to ensure PDN stability~\hp{\mbox{\cite{fayneh20164,jeong2012mapg,haj2021ichannels,kahng2013many,akl2009effective,kahng2012tap}}}.
We further discuss this in \autoref{sec:control_in_rush}.
The remaining capabilities, such as clock-gating, event detection (interrupts, snoops) are all commonly supported in state-of-the-art SoCs.
The C6A controller is implemented using a simple finite-state-machine (FSM) within the core's PMA, which resides in the uncore\mbox{~\cite{rotem2012power}} and controls clock gating/un-gating, save/restore signals, and L1/L2 entry-to/exit-from sleep-mode.
 The $C6A$ snoop flow reuses the existing snoop handling mechanisms of the $C1$ state (shown in Fig.~\mbox{\ref{fig:c1_c6_c_state_flow}}).

\noindent \textbf{\hyy{Power overhead.}}
\tech implements the C6A controller as an FSM within the PMA.
Based on a comparable power-management flow implemented in~\cite{haj2020techniques}, we estimate that the C6A controller adds approximately $5mW$ to the PMA power.

\noindent \textbf{\hyy{Performance overhead.}} \hj{The additional control circuit we add to the PMA has no direct} impact on CPU core's performance. \hj{The performance overhead is mainly due to the new features the PMA controls (described in \autoref{sec:perf_penalty} and \autoref{sec:ccsm}).} 

\noindent \textbf{\hyy{Area overhead.}}
Based on a comparable power-management flow implemented in\mbox{~\cite{haj2020techniques}}, we estimate the additional area to implement the C6A controller to be up to $5\%$ of the core PMA area.

\subsubsection{Core PLL and FIVR}
\hj{\tech keeps the PLL and FIVR powered on in C6A/C6AE states. Next we describe only the power overhead of the PLL and FIVR as there is no additional performance or area overhead as compared to the baseline.}

\noindent \textbf{\hyy{Power overhead.}}
We estimate the C6A idle power needed by \tech to keep the PLL on and locked while accounting for voltage regulator inefficiencies.
The Skylake core uses an ADPLL and a FIVR~\cite{fayneh20164,Skylake_die_server}.
The ADPLL consumes $7$mW (fixed across core voltage/frequency \jk{levels}~\cite{fayneh20164}).
The FIVR presents dynamic efficiency loss due to conduction and switching inefficiency~\cite{lakkas2016mosfet}, and static efficiency loss due to power consumption of the control and feedback circuits~\cite{lakkas2016mosfet,haj2020flexwatts,nalamalpu2015broadwell}.
The static loss still applies when the FIVR output is \textit{$0V$}.
The FIVR static loss accounts for ${\sim}100$mW per core~\cite{haj2019comprehensive,haj2020flexwatts,asyabi2020peafowl}.
The FIVR efficiency at light load is about $80\%$ (excluding the static power looses)~\cite{haj2020flexwatts,haj2019comprehensive,radhakrishnan2021power}.   


%
%
\newcommand\firstcolsize[0]{3cm}
\newcommand\markerA[0]{\textsuperscript{\,$\alpha$}}
\newcommand\markerB[0]{\textsuperscript{\,$\beta$}}
\newcommand\markerC[0]{\textsuperscript{\,$\gamma$}}
\newcommand\markerD[0]{\textsuperscript{\,$\delta$}}
\newcommand\markerE[0]{\textsuperscript{\,$\epsilon$}}
\newcommand\markerF[0]{\textsuperscript{\,$\zeta$}}
\newcommand\markerG[0]{\textsuperscript{\,$\eta$}}

\begin{table*}[tb]
\small
\centering
\caption{Area and power requirements to implement \tech in \jk{an Intel} Skylake-like core.}
\label{tab:area_poweroverhead}
\resizebox{1.0\linewidth}{!}{%
\def\arraystretch{1.2}
\begin{tabular}{p{\firstcolsize}p{4.9cm}p{4.6cm}p{2.5cm}p{2.5cm}}
\multicolumn{1}{c}{\bf Component} & \multicolumn{1}{c}{\bf Sub-Component} & \multicolumn{1}{c}{\bf Area Requirement} & \multicolumn{1}{c}{\bf C6A Power} & \multicolumn{1}{c}{\bf C6AE Power} \\
\cmidrule(lr){1-1} \cmidrule(lr){2-2} \cmidrule(lr){3-3} \cmidrule(lr){4-4} \cmidrule(lr){5-5}
\multirow{4}{\firstcolsize}{\textbf{Units' Fast Power-Gating (UFPG)}} & {Unit} power-gates (${\sim}70\%$ of the core) & $2$\,--\,$6\%$ of power-gated area & ${\sim}30\,$--\,$50$mW\markerA & ${\sim}18\,$--\,$30mW$\markerA \\ 
 & Ungated context registers & ${<}1\%$ of ungated context registers & \multirow{3}{*}{${\sim}2$mW\markerB} & \multirow{3}{*}{${\sim}1$mW\markerB} \\
 & State retention power-gates (SRPG) & ${<}1\%$  of gated unit area & \\
 & Ungated context SRAM & ${<}1\%$ of SRAM area & \\ 
\cmidrule(lr){1-1} \cmidrule(lr){2-2} \cmidrule(lr){3-3} \cmidrule(lr){4-4} \cmidrule(lr){5-5}
\multirow{2}{\firstcolsize}{\textbf{Cache Coherence
\& Sleep Mode (CCSM)}} & L1/L2 caches {in} sleep-mode & $2\,$--\,$6\%$ of private cache area & $55$mW\markerC & $40$mW\markerC\markerD  \\
 & The rest of the memory subsystem & ${<}1\%$ of the ungated units & $55$mW\markerC & $33$mW\markerC \\ 
\cmidrule(lr){1-1} \cmidrule(lr){2-2} \cmidrule(lr){3-3} \cmidrule(lr){4-4} \cmidrule(lr){5-5}
\textbf{PMA Flow} & Implemented in the uncore~\cite{rotem2012power} & ${<}5\%$ of core PMA~\cite{haj2020techniques} & $5$mW\markerE & $5$mW\markerE \\ 
\cmidrule(lr){1-1} \cmidrule(lr){2-2} \cmidrule(lr){3-3} \cmidrule(lr){4-4} \cmidrule(lr){5-5}
\multirow{3}{\firstcolsize}{\textbf{Core ADPLL \& FIVR}}& ADPLL & $0\%$ & $7$mW~\cite{fayneh20164} & $7$mW~\cite{fayneh20164} \\
 & Core FIVR inefficiency & $0\%$ & $36$mW\,--\,$41$mW\markerF  & $23$mW\,--\,$27$mW\markerF  \\ 
& FIVR static losses & $0\%$ &  $100$mW\markerG &  $100$mW\markerG \\
\cmidrule(lr){1-1} \cmidrule(lr){2-2} \cmidrule(lr){3-3} \cmidrule(lr){4-4} \cmidrule(lr){5-5}
\textbf{Overall} & & {\boldmath{$3$}\,--\,\boldmath{$7\%$} \bf{of the core area}} & {\boldmath{$290$\textbf{\,--\,}\boldmath{$315$}\textbf{mW}}} & {\boldmath{$227$}\textbf{\,--\,}\boldmath{$243$}\textbf{mW}} 
\end{tabular}%
}
\begin{flushleft}
\markerA Assuming $3$\,--\,$5\%$~\cite{flynn2007low} of leakage power~$\approx$ C1 power. \quad
\markerB Power of the ${\sim}8KB$ context~\cite{haj2020techniques}. \quad
\markerC L1+L2 size is ${\sim}1.1MB$; power from~\cite{chen201322nm_ppt} \hy{in $22$nm} scaled to $14$nm based on~\cite{shahidi2018chip}. \quad
\markerD Higher \jk{sleep-transistor} efficiency at $V_{min}$ (C6AE) \hy{\mbox{\cite{luria2016dual,huang2016fully,haj2020flexwatts}}}. \quad
\markerE Based on scaled wake-up logic power from~\cite{haj2020techniques}. \quad
\markerF Assuming $80\%$ FIVR efficiency in light load~\hy{\mbox{\cite{haj2020flexwatts,haj2019comprehensive,radhakrishnan2021power}}}.  \quad 
\markerG FIVR static losses~\hy{\mbox{\cite{haj2019comprehensive,haj2020flexwatts,asyabi2020peafowl}}} in C6 state are ${\sim}100$mW~\cite{nalamalpu2015broadwell}.
\end{flushleft}
\end{table*}

\subsection{C6A and C6AE Latency}\label{sec:c6a_lat}
We estimate the overall transition time (i.e., entry followed by direct exit) for the C6A and C6AE states of \tech to be ${<}100$ns. This is \ho{three orders} of magnitude faster than the ${>}100${\textmu}s \jk{latency} of C6.
\jk{Next,} we explain in detail this estimation by referring to the flow in~\autoref{fig:c6a_c_state_flows}.

\subsubsection{C6A and C6AE Entry Latency} 
Clock-gating all \ho{domains} and keeping the PLL ON \jk{(\circledb{1} in \autoref{fig:c6a_c_state_flows})} typically takes $1$\,--\,$2$ cycles in an optimized clock distribution system~\cite{el2011clocking,shamanna2010scalable}.
Transitioning to $Pn$ (required for $C6AE$) happens with a non-blocking parallel \hj{DVFS (i.e., P-state)} flow that \hj{can take few tens} of microseconds, depending on the power management architecture~\cite{gendler2021dvfs}.
Since \tech keeps \jk{the} context in-place; saving the context to power-gate the core units \circledb{2} only requires asserting the \texttt{Ret} signal followed by deasserting the \texttt{Pwr} signal, as shown in \autoref{fig:state_retention}(c).
We estimate this process to take $3$--$4$ cycles.
Finally, placing the L1/L2 caches in sleep-mode and clock-gating them \circledb{3} takes $1$--$3$ cycles. 
Hence, the overall entry flow takes $<10$ cycles, \jk{i.e.,} less than $<20ns$ with a power management controller clocked at $500MHz$.%
\footnote{Typically, a power management controller of a modern SoC operates at a frequency of several \jk{hundred} MHz (e.g., 500MHz \cite{peterson2019fully}) to handle \jk{nanosecond-scale} events, such as \textit{di/dt} noise\hp{\mbox{\cite{fayneh20164}\cite[Sec. 5]{haj2021ichannels}}}.} 

\subsubsection{C6A and C6AE Exit Latency} 
Clock-ungating the L1/L2 caches and exiting sleep-mode \circledb{4} takes $2$ cycles~\cite{huang2013energy}.
Power-ungating the core units \circledb{5} takes ${<}70ns$ (discussed in \autoref{sec:control_in_rush}) and restoring the core context (i.e., deasserting the \texttt{Ret} signal after restoring power) takes $1$ cycle.
Finally, clock-ungating all domains \circledb{6} typically takes $1$--$2$ cycles. 
Hence, the overall exit flow takes ${\sim}5$ clock \jk{cycles} + ${<}70ns$, equivalent to ${<}80ns$ when using a $500MHz$ clock.  

\subsubsection{C6A and C6AE Snoop Handling} 
Snoop handling latency in C6A(C6AE) is similar to that \jk{in} C1(C1E).
Specifically, clock-ungating the L1/L2 caches and exiting sleep-mode \circleda{a} takes $2$ cycles.
In the first cycle, the flow ungates the clock; in the second cycle, the snoop requests simultaneously 1) \jk{access} the cache tags (power-ungated), and 2) wake-up the cache data array~\cite{huang2013energy,chen201322nm,rusu20145,rusu201422}.
Placing the L1/L2 in sleep-mode and clock-gating L1/L2 caches \circleda{c} after servicing the snoop traffic \circleda{b} takes $1$\,--\,$3$ cycles.

\begin{table*}[ht]
\centering
\caption{\hg{Comparison of Core Power-gating Schemes}}
\label{tab:pg_schemes_compare}
\resizebox{0.8\linewidth}{!}{%
\begin{tabular}{lllll}
\textbf{Technique} & \textbf{Core Type} & \textbf{Power-gating Trigger} & \textbf{Power-gated Blocks} & \textbf{Wake-up Overhead} \\ 
\cmidrule(lr){1-1} \cmidrule(lr){2-2} \cmidrule(lr){3-3} \cmidrule(lr){4-4} \cmidrule(lr){5-5}
\cite{roy2010state}             & In-order  CPU      & Cache miss                 & Register file                                                         & 5 cycles                                                            \\ 
\cite{jeong2012mapg}            & In-order CPU       & Cache miss                 & Core                                                                  & 10ns                                                                \\ 
\cite{hu2004microarchitectural} & OoO CPU   & Execution unit idle        & Execution units             & 9 cycles                                                            \\ 
\cite{battle2012flexible}       & OoO CPU   & Register file bank idle  & Register file bank          & 17 cycles                                                           \\ 
\cite{jeon2015gpu}              & GPU                                                          & Register subarray unused & Register subarray           & 10 cycles                                                           \\ 
\cite{haj2021ichannels}         & OoO CPU   & AVX execution unit idle    & Intel AVX  execution unit          & $\sim$10-15ns                                                                \\ 
\tech (This work)                       & OoO CPU   & Core idle                                                            & Most of  core units             & $\sim$70ns                                                               \\ 
\end{tabular}
}
\end{table*}

\subsection{Staggered Unit Wake-up}
\label{sec:control_in_rush}
As discussed in \autoref{sec:bg}, \jk{rapid} wake-up of a power-gated domain can result in a sudden increase in current demand (in-rush current)~\cite{chadha2013architectural,usami2009design,agarwal2006power,abba2014improved}, \jk{which} can damage a chip. \jk{Intel} Skylake core's AVX power-gating mitigates this by staggering the AVX un-gating over ${\sim}15ns$~\hp{\cite{fayneh20164}\cite[Sec. 5]{haj2021ichannels}}. 
\tech can exacerbate in-rush current effects, since during C6A\,/\,C6AE exit it wakes up a power-gated domain (i.e., \UFPG, the red shaded area in Fig. \ref{fig:skx_ufpg}) that has approximately $4.5\times$ \hjj{the} area and capacitance \hjj{of} the AVX units~\cite{haj2016fine}. \hoo{We avoid this issue by dividing the UFPG area into five zones, each with a local power-gate controller (as in~\mbox{\autoref{fig:staggered_pg})}.
Each of the five 
controllers has a zone sleep signal (\texttt{$SlpZone_i$}) that is controlled by the core PMA.
The PMA sequentially wakes up the five domains using the \texttt{$SlpZone_i$} signals.
Since each of the five zones has a smaller area than the AVX power-gated units, staggering the wake up of each zone over ${\leq}15ns$ (i.e., same as AVX units) keeps the in-rush current within limits~\mbox{\cite{chadha2013architectural,jeong2012mapg,agarwal2006power}}.
Hence, waking up all five domains\hj{, which have ${\sim}4.5\times$ \hjj{the} area and capacitance \hjj{of} the AVX units,} takes approximately ${<}70$ns \hj{($67.5=4.5\times15$ns)}.}

\hg{Several prior works propose nanosecond-scale staggered \jk{power-gate} wake-up for different units\jk{,} including the entire core. Table \mbox{\ref{tab:pg_schemes_compare}} summarizes some of these works.}

\subsection{Design Complexity and Effort}
\tech techniques involve non-negligible front-end and back-end design complexity and effort. 
Even though medium-grained power-gates of \UFPG are less invasive than \jk{fine-grained} power-gating, they still require significant back-end (e.g., power-delivery, floor-planning, place and route) effort.
Moreover, the \CCSM and the C6A/C6AE  \emph{control flows} \jk{require} careful pre-silicon  verification to ensure that all the hardware flows (Fig. \ref{fig:c6a_c_state_flows}) operate according to the architectural specification. The complexity and effort can be even worse if a processor vendor chooses to have two separate designs for client and server to \jk{remove} \tech's overhead \jk{from client systems}. 

Nonetheless, \tech complexity and effort \jk{are} comparable to recent techniques implemented in modern CPUs to increase energy-efficiency\hj{, such as hybrid cores \cite{apple_m1,rotem2021alder}}. Therefore, once there is a strong demand from customers and/or pressure from competitors, CPU vendors \jk{can implement an architecture similar} to \tech to significantly increase \jk{server} energy efficiency. 

\subsection{\tech Benefits to  AMD Processors}\label{sec:amd_benefits}
Despite having a hierarchical design\hj{, which uses chiplet dies (CCDs) and clusters of CPU cores (CCXs) within each CCD \cite{E1},} AMD's modern server processors still suffer from the same issues that we pointed out in \hjj{\autoref{sec:motivation}}. Since the latency to enter/exit \hjj{a} \hj{CPU} core deep idle state is tens \hjjj{or hundreds} of microseconds \cite{E1}, server vendors recommend disabling the deep idle C-state (Global C-State Control in BIOS) in AMD EPYC Rome/Milan-based servers to reduce performance impact \cite{E2, E3, E4}. Therefore, despite the capability to place individual idle cores (or even \hj{a cluster of cores}) in a deep low-power state, this capability is typically disabled in AMD servers running latency-critical applications, which significantly increases the energy consumption of these servers \cite{E2}. \tech can mitigate this issue by providing a low-power C-state with nanosecond-scale transition latency.

%% file: _06_methodology.tex
\section{Experimental Methodology}
\label{sec:methodology}

\subsection{Workloads and Experimental Setup}
We evaluate \tech using three latency-critical applications: \emph{Memcached}, \emph{Apache Kafka}, and \emph{MySQL}.
Memcached~\cite{memcached} is a lightweight key-value store that is widely deployed as a distributed caching service to accelerate user-facing applications with stringent latency requirements~\cite{nishtala2013memcached, yang2020twemcache, pymemcache}.
Memcached has been the focus of numerous studies\hjj{~\cite{gan2021sage,lim2012system,lim2013memcached,xu2014memcached,leverich2014mutilate,prekas2017zygos}}, including efforts to provide low microsecond-scale tail latency\hjj{~\cite{mirhosseini2020q,chow2014mystery,nishtala2017hipster,prekas2015energy,mirhosseini2019express,sriraman2018mutune,hsu2017reining,hsu2015adrenaline,sriraman2018mu,nishtala2013memcached, jialin2014memcached, asyabi2020peafowl}}.
Kafka~\cite{kreps2011kafka} is a real-time event streaming platform that is used to power event-driven microservices and stream processing applications.
MySQL~\cite{mysql} is a widely\hj{-}used relational database management system.



Our baseline server is equipped with two Intel Xeon Silver 4114 \cite{Skylake_4114} Skylake-based processors running at a base frequency of 2.2 GHz (minimum frequency of 0.8 GHz and maximum Turbo Boost frequency of 3 GHz), with 10 physical cores for a total of 20 hyper-threads, and with 192 GB DDR4 DRAM.
We use a cluster of \hj{six} server machines to run the Memcached, Kafka, and MySQL services and corresponding workload clients. 
For evaluating Memcached, we run a single Memcached server process on one of the \hj{server} machines and run a modified version of the Mutilate load generator~\cite{leverich2014mutilate} for Memcached on \hjj{the remaining} \hj{five server} machines. 
We configure the load generator to recreate the ETC workload from Facebook \cite{atikoglu2012workload}, using one master and four workload-generator clients, each running on a separate \hj{server} machine. 
For evaluating Apache Kafka, we run a single Kafka server process on one \hj{server} machine and run the ConsumerPerformance and ProducerPerformance Kafka tools on another \hj{server machine}.
For evaluating MySQL, we run a single MySQL server process on one \hj{server} machine and run the \texttt{sysbench} benchmarking tool with the OLTP test profile on another \hj{server machine}. 
In all cases, we pin the processes to specific cores to minimize the impact of the OS scheduler.



\subsection{Power and Performance Model}\label{sec:power-model}
We model the average power of a core using the fraction of time spent at each unique C-state and its corresponding power consumption. 
Similar to \hj{prior} works~\cite{kasture2015rubik,lo2014towards,kanev2014tradeoffs,chou2016dynsleep,fan2007power,chou2019mudpm,asyabi2020peafowl,mirhosseini2019enhancing,zhan2016carb}, we focus on CPU power which is the single largest contributor to server power \cite{jin2020review,vasques2019review}. \hp{\hj{Next, we} describe our models for the baseline and \mbox{\tech}.}

\noindent \textbf{\hp{Modeling the Baseline CPU Core.}} \hjj{Our} analytical power model estimates the average CPU core power ($Avg_{P}$) for a workload, \hoo{assuming P-state\hj{s}/Turbo are disabled,} as follows: 
\begin{align}
  \resizebox{0.55\hsize}{!}{$Avg_{P} = \sum_{i \in \{0, 1, 1E, 6\}} P_{C_i} \times R_{C_i}$}
\end{align}
\noindent
$P_{C_i}$ denotes the core power in state $Ci$ {(reported in Table \ref{tab:c-states})}. 
$R_{C_i}$ denotes the residency at $Ci$, i.e., 
the percentage of the total time the system spends at state $Ci$. 
We obtain C-state residency and number of transitions  using processor's residency reporting counters~\cite{intel_skl_dev}. \hj{When executing our workloads,} \hjj{we} use the RAPL interface \cite{HDC_intel} to measure power consumption. 

\noindent \textbf{\hp{Modeling the \mbox{\tech} CPU Core.}} 
We model the power consumption 
of the CPU core enhanced with the two new C-states of \tech (i.e., $C6A$ and $C6AE$) using 1) measured data from our baseline power model; C-state residency is scaled using our performance model (more details below), and 2) estimated power of the $C6A$ and $C6AE$, as summarized in Table \ref{tab:area_poweroverhead}. $C6A$ and $C6AE$ C-states of \tech replace the $C1$ and $C1E$, respectively, as follows:
\begin{align}
  \resizebox{0.6\hsize}{!}{$Avg_{P} = \sum_{i \in \{0, 6A, 6AE, 6\}} P_{C_i} \times R_{C_i}$}
\end{align}
\noindent
Therefore, for a given workload, we perform the following steps. 
1) We obtain the power and residency of each core C-state from the baseline. We scale the C-state residency taking into account i) how the small core frequency degradation incurred due to the power-gates (Sec.~\ref{sec:perf_penalty}) affects performance by considering a workload's frequency scalability\footnote{We define \hjj{the} frequency scalability of a workload \hjj{as} \hj{the change in its performance} with unit change in frequency, as in~\cite{yasin2017performance,haj2015doee,gendler2021dvfs,yasin2019metric}.}
and ii) the \hj{higher} $C6A$/$C6AE$ transition latency (i.e., $100ns$; Sec.~\ref{sec:c6a_lat}) compared to $C1$/$C1E$.
2) We replace $C1$/$C1E$ C-state residency (i.e., $R_{C_1}$/$R_{C_{1E}}$) with $C6A$/$C6AE$ C-state residency (i.e.,$R_{C_{6A}}$/$R_{C_{6AE}}$).  
3) We replace $C1$/$C1E$ power consumption (i.e., $P_{C_1}$/$P_{C_{1E}}$) with $C6A$/$C6AE$ \emph{{estimated}} power consumption (i.e., $P_{C_{6A}}$/$P_{C_{6AE}}$, in Table \ref{tab:area_poweroverhead}).  We {plug in} the new values 
to estimate the average \tech CPU core power. 

\hoo{To estimate \mbox{\tech's} average power for workloads with Turbo enabled (and P-state disabled), we use the following Equation:}
\begin{align}
& \resizebox{0.9\hsize}{!}{$AvgP_{savings} = R_{C_1} \times (P_{C_1} - P_{C_{6A}}) + R_{C_{1E}} \times (P_{C_{1E}} - P_{C_{6AE}})$} \nonumber \\
& \resizebox{0.73\hsize}{!}{$AvgP_{savings\%} = (AvgP_{savings} / AvgP_{baseline}) \times 100$} 
\label{eq:ideal-savings-turbo}
\end{align}
%
\hoo{where we measure $AvgP_{baseline}$ using RAPL. Doing so allows our model to take into account power consumption variation during $C0$ active state due to Turbo transitions.}



\subsection{Power Model Validation}
The power consumption at each processor C-state and frequency step (i.e., P1 and Pn) is collected from measurements of real systems based on the Intel Skylake CPU~\cite{tam2018skylake}\hjj{, which} is shown in Table \ref{tab:c-states}.
To validate our model, we run \hp{four} representative server workloads: SPECpower \cite{lange2009identifying}, Nginx \cite{nginx}, Spark \cite{spark}, and Hive \cite{hive} at multiple CPU utilization levels.
We measure average  power \hj{consumption} and collect core C-states residencies and transitions for each run. We use our analytical power model to estimate the average power consumption of these workloads. Then, we compare the measured vs. estimated average power. We find that the accuracy of our 
model is  \hp{$96.1\%$ / $95.2\%$ /  $94.4\%$ / $94.9\%$ for SPECpower / Nginx  / Spark / Hive  workloads, respectively}.

%% file: _07_results.tex
\section{Evaluation}
\label{sec:evaluation}


We first evaluate \tech 
using the Memcached~\cite{memcached} \hj{workload}. In Sec. \ref{sec:other_workloads} we evaluate \tech with two more workloads.  

\subsection{Power Savings and Overhead at Varying Load Levels}
\begin{figure}[ht]
\centering
\includegraphics[trim=0.6cm 0.75cm 0.6cm 0.8cm, clip=true,width=0.99\linewidth,keepaspectratio]{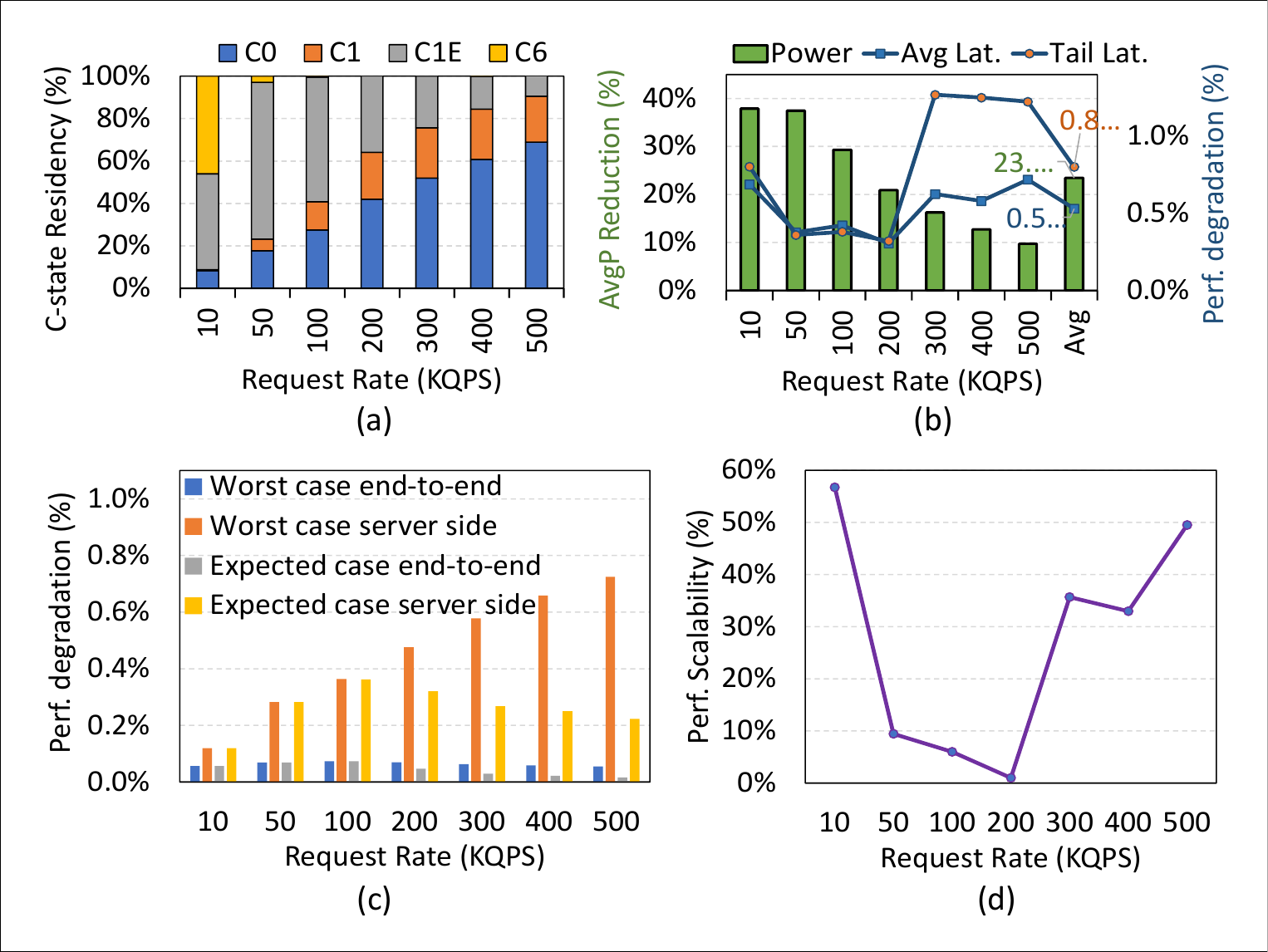}
\caption{%
\rG{Comparison of \tech against the baseline configuration (P-state disabled, Turbo enabled, C-state enabled) with varying request rates. (a) Residency of the baseline system at different C-states. (b) \tech core average power (AvgP) reduction and average/tail latency degradation when replacing $C1$/$C1E$ with $C6A$/$C6AE$. (c) Average response time degradation. (d) Performance scalability when increasing frequency from $2$\,GHz to $2.2$\,GHz.}}
\label{fig:res_baseline_compare}
\end{figure}

Fig.~\ref{fig:res_baseline_compare} shows how \tech affects power consumption and request latency against the baseline with \hoo{P-states disabled (i.e., frequency is set to base frequency, $P1$)} and C-states and Turbo enabled. We expect \tech to achieve significant power savings, thanks to the lower power of $C6A$/$C6AE$ compared to $C1$/$C1E$, while having a small impact on request latency because of the \hj{larger} transition time (${\sim}100ns$) and the ${\sim}1\%$ frequency loss (Sec.~\ref{sec:perf_penalty}).
While the \hj{larger} transition time impacts each C-state transition, the impact of the frequency loss depends on the sensitivity of the workload to the core frequency reduction.

Fig.~\ref{fig:res_baseline_compare}(a) shows the residency of the system in each C-state: as expected, idle time progressively \hj{reduces} as load (queries per second -- QPS) increases.
Therefore, we expect \tech to have larger power savings and lower impact on performance at low load.
Indeed, Fig.~\ref{fig:res_baseline_compare}(b) shows that \tech reduces the average power consumption by up to $38\%$ at low load, with less than $1\%$ impact on both average and tail latency.
At high load, \tech still provides $10\%$ power savings, with less than $1.3\%$ impact on tail latency.
For reference, Fig.~\ref{fig:res_baseline_compare}(d) shows the \hj{performance} scalability of {\tt Memcached} to increasing core frequency from $2$\,GHz to $2.2$\,GHz (Sec. \ref{sec:power-model}).

Fig.~\ref{fig:res_baseline_compare}(c) further analyzes the impact of \tech on average response time.
We consider end-to-end (including network latency measured at 117us) and server-side response time for two cases: the worst case, where we assume a C-state transition for each query and the expected case, with the actual C-state transitions observed in the baseline.
As expected, the gap between the worst and the expected case is larger at high load, since multiple queries are serviced within the same active period.
We also observe that the degradation of the end-to-end response time (i.e., by client) is negligible because the (non-changing) network latency dominates the overall response time.

Google states in their paper that discusses latency-critical applications \mbox{\cite{lo2014towards}}: ``Modern servers are not energy proportional: they operate at \emph{peak} energy efficiency when they are fully utilized, but have much lower efficiencies at lower utilizations''. The utilization of servers running latency-critical applications is only $5\%$--$25\%$ to meet target tail latency requirements, as reported by multiple works from academia and industry \mbox{\cite{lo2014towards,B1,B2,B3,B4}}. For example, recently, Alibaba reported that the utilization of servers running latency-critical applications is typically 10\% \mbox{\cite{B4}}. Therefore, our AW proposal addresses the more inefficient aspect of modern servers: running latency-critical microservice\hj{-}based applications at low utilization.

We conclude that \tech significantly \hj{reduces} core average power consumption of the {\tt Memcached} service across \hj{various} load levels with minimal performance overhead over the baseline when disabling P-states and enabling Turbo.

\subsection{Commonly-Used Configurations}

Server vendors 
provide recommended system configurations~\cite{cisco_cstates,dell_cstates,lenovo_cstates}, such as disabling certain C-states to increase system performance or disabling Turbo Boost \hj{\cite{rotem2013power,rotem2012power}} to reduce power consumption.
We analyze three common configurations by modifying our baseline (which has P-states disabled and Turbo and C-states enabled) by successively disabling Turbo, $C6$, and $C1E$.
Before analyzing the impact of \tech on these three tuned configurations, we study them individually.

\begin{figure}[t]
\centering
\includegraphics[trim=0.6cm 0.75cm 0.6cm 0.8cm, clip=true,width=0.99\linewidth,keepaspectratio]{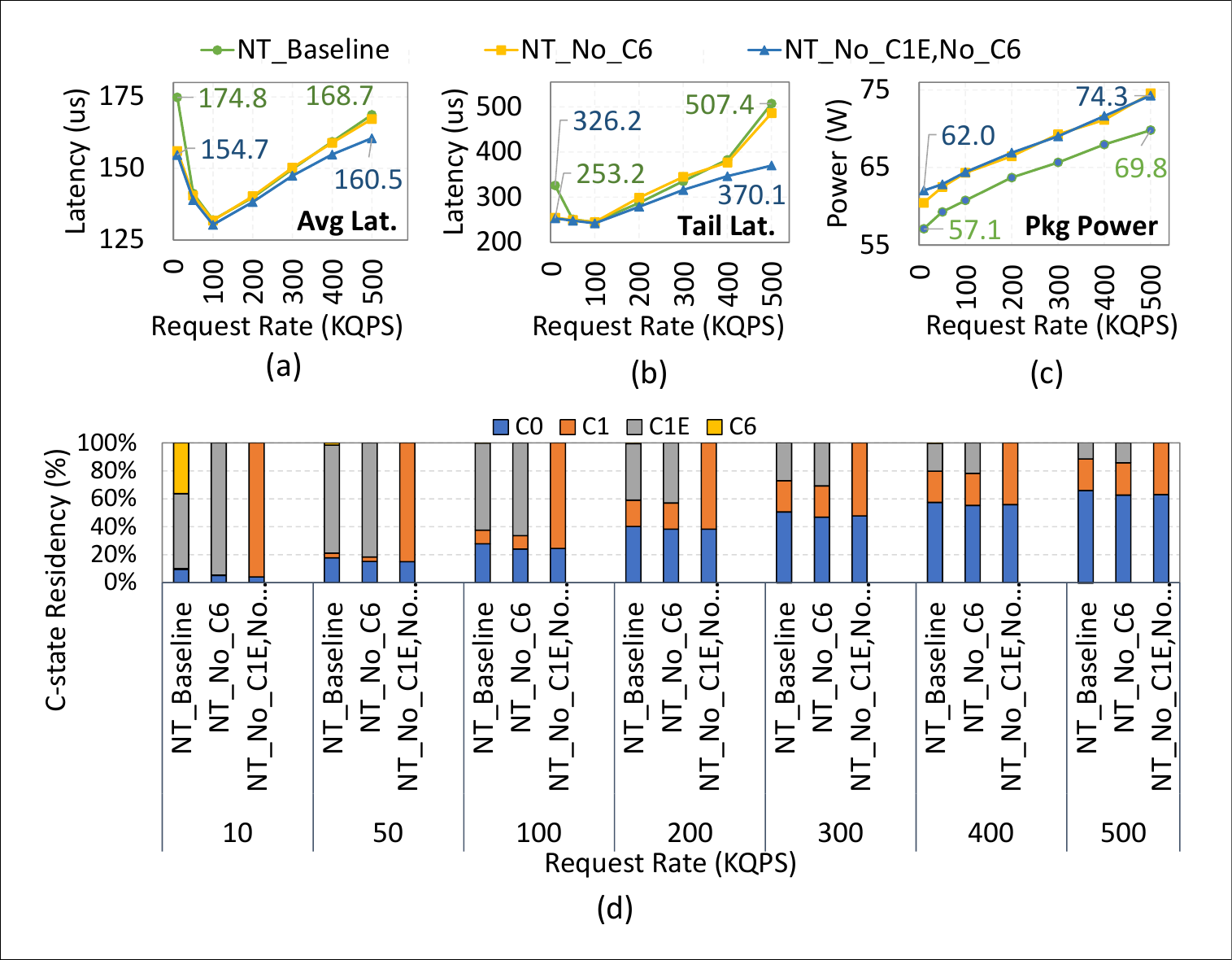}
\caption{%
Analysis of three variants of the baseline configuration (\texttt{NT\_Baseline} disables Turbo, \texttt{NT\_No\_C6} disables Turbo and $C6$, \texttt{NT\_No\_C6,No\_C1E}) disables, Turbo, $C6$, and $C1E$) on (a) Average latency, (b) Tail latency, (c) Package power \hj{consumption}, (d) C-state residency at increasing load level. All configurations have P-states disabled. 
}
\label{fig:res_multi_cfg}
\end{figure}

Fig.~\ref{fig:res_multi_cfg} reports latency (average and tail), package power \hj{consumption}, and C-state residency for the three tuned configurations.
We observe that \texttt{NT\_No\_C6,No\_C1E} has the lowest average and tail latency, but also the highest average power across the entire range of request rates.
At $500$\,KQPS, this configuration has ${\sim}5\%$ and ${\sim}27\%$ lower average and tail latency, respectively, but also has ${\sim}7\%$ higher average package power than the other two configurations.  
Latency improves because disabling $C1E$ reduces the long transition overhead of $C1E$ (i.e., $10{\mu}s$, shown in Table~\ref{tab:c-states}), in contrast to the other two configurations that spend significant time in $C1E$, as shown in Fig.~\ref{fig:res_multi_cfg}(d).\footnote{\hgg{We explain the irregular performance trends in Fig. \ref{fig:res_multi_cfg} with the following observations: at low load, the cores enter $C6$ state, which increases the query response latency by the latency overhead required to transition from $C6$ to the $C0$ active state. As the load increases, two phenomena occur: 1) fewer transitions to $C6$ state and 2) queueing effects. In the mid-range load, there are fewer queueing effects and \hjjj{fewer} transitions to $C6$ \hj{and} that's why the \hj{latency} is \hj{lowest}, whereas at high load, even though \hjjj{the} transitions to $C6$ \hjjj{are the lowest}, the latency is dominated by queueing effects. 
}\label{footnote:eval_expl}}
However, disabling $C1E$ also increases average power \hj{consumption} because, as shown in Fig.~\ref{fig:res_multi_cfg}(d), the core now spends more time in $C1$ (the shallower C-state), which has ${\sim}63\%$ higher power than $C1E$, as shown in Table \ref{tab:c-states}.
This analysis shows that a new C-state that consumes similar (or lower) power to $C1E$ but with a transition time that is close to $C1$ can provide a low average and tail latency with reduced power consumption. Next, we show that our newly proposed C-state, $C6A$, achieves this balance.

\begin{figure}[ht]
\centering
\includegraphics[trim=0.6cm 0.75cm 0.6cm 0.8cm, clip=true,width=0.99\linewidth,keepaspectratio]{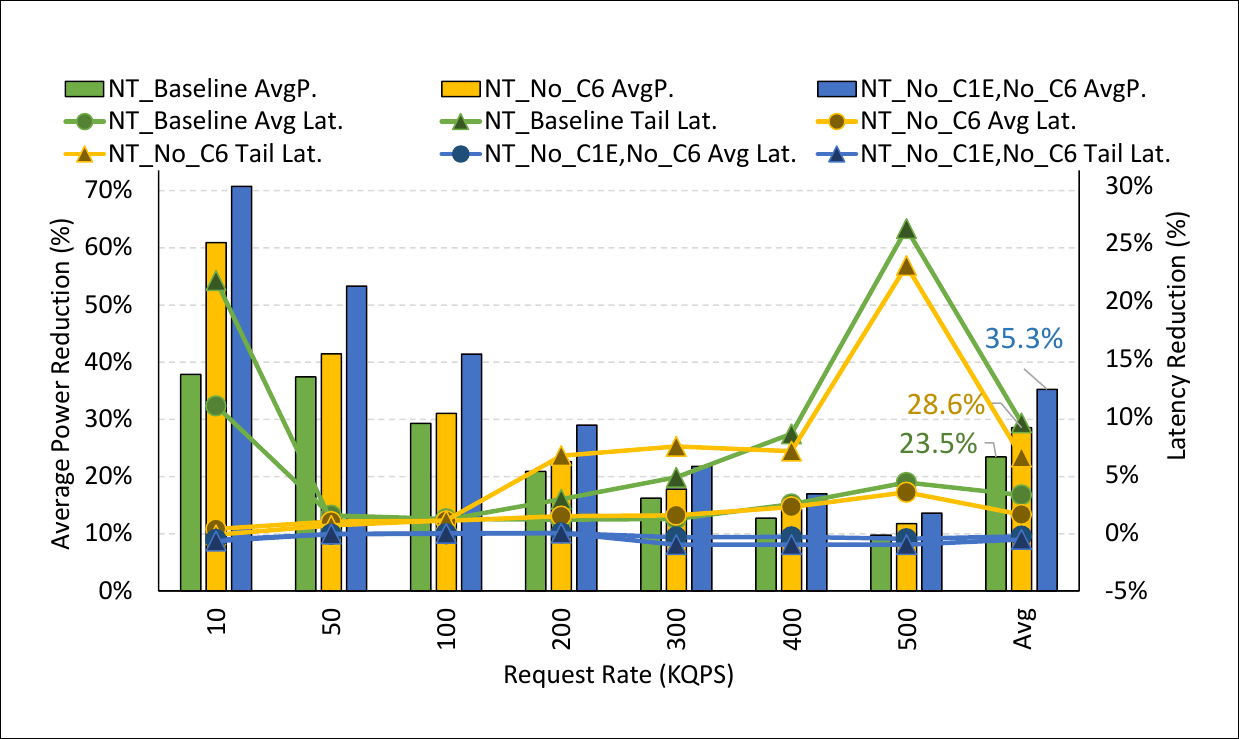}
\caption{\hj{Power reduction and average/tail latency reduction} of \tech\hjj{~over} baseline (P-state disabled,  C-state enabled) with Turbo disabled (\texttt{NT\_Baseline}), Turbo and $C6$ disabled (\texttt{NT\_No\_C6}), and Turbo, $C6$ and $C1E$ disabled (\texttt{NT\_No\_C6,No\_C1E}) \hj{across}  different request rates (QPS).}
\label{fig:res_multi_cfg_AW}
\end{figure}

Fig.~\ref{fig:res_multi_cfg_AW} \hj{shows the power reduction and performance (tail and average latency) improvement of \tech over} the three tuned configurations. \hj{We observe that \tech significantly reduces power consumption against all three tuned configurations.}
The reason is that, in these workloads \tech replaces the time that other configurations spend in $C1$/$C1E$ with \hj{the} $C6A$/$C6AE$ \hj{C-states}, which \hj{have} much 
lower power.
\hj{Second}, \tech~\hjj{reduces average/tail latency} \hjj{by up to $5\%/26\%$ and $4\%/24\%$} compared to \texttt{NT\_Baseline} and \texttt{NT\_No\_C6}\hjj{, respectively,} while only degrading performance by less than $1\%$ compared to \texttt{NT\_No\_C6,No\_C1E}.
Based on this analysis, we conclude that \tech provides average and tail latencies comparable to \hj{or better than} the tuned configurations, while reducing power \hj{consumption} significantly.

\subsection{Analysis of Turbo  Performance Improvement}
To maximize performance, server vendors recommend to enable Turbo for better burst performance and disable $C6$ and $C1E$ to avoid their transition latency~\cite{cisco_cstates,dell_cstates,lenovo_cstates}.
However, server vendors also note that disabling $C1E$ can hamper performance since the processor is kept at high power, thereby not gaining enough thermal capacitance needed during Turbo Boost periods~\cite{lenovo_cstates,rotem2011power,rotem2012power,rotem2013power,rotem2015intel}.
Therefore, current C-state architectures cannot benefit from
removing the $C1E$ transition overhead and enabling Turbo.
In this section, we demonstrate that \tech achieves high Turbo performance while eliminating the $C1E$ performance overhead. 

\begin{figure}[ht]
\centering
\includegraphics[trim=0.6cm 0.6cm 0.6cm 0.6cm, clip=true,width=0.88\linewidth,keepaspectratio]{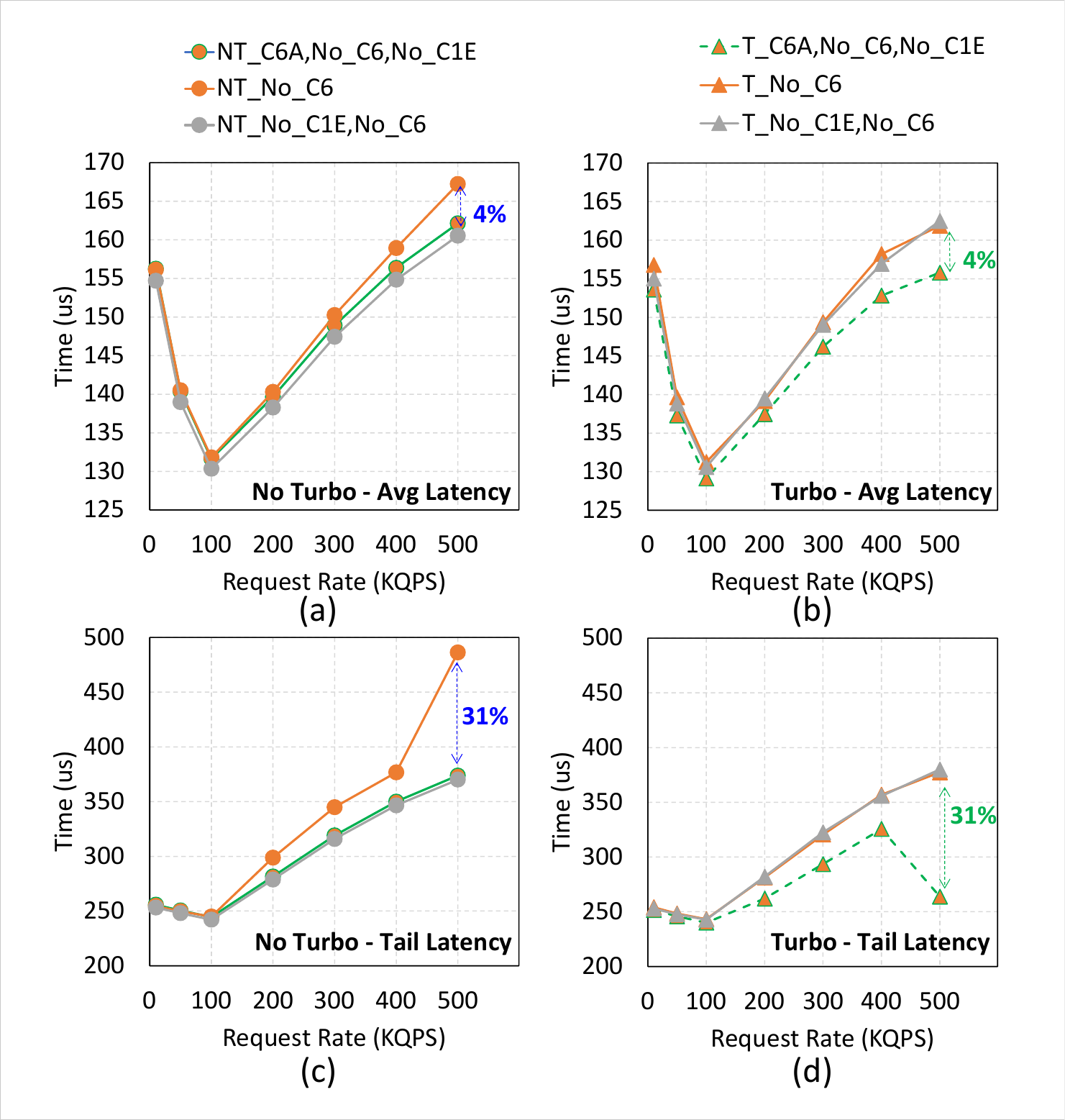}
\caption{
Average and tail latency at different request rates (QPS) for four configurations that show the effect of idle power state on Turbo performance:
Turbo/No-Turbo \& $C6$ disabled (\texttt{T\_No\_C6}/\texttt{NT\_No\_C6}), Turbo/No-Turbo \& $C6$ \& $C1E$ disabled (\texttt{T\_No\_C6,No\_C1E}/\texttt{NT\_No\_C6,No\_C1E}), compared with \tech: \hy{Turbo/No-Turbo \& $C6A$ enabled  \& $C6$ \& $C1E$ disabled (\texttt{T\_C6A,No\_C6,No\_C1E}/\texttt{NT\_C6A,No\_C6,No\_C1E})}. 
}
\label{fig:res_turbo}
\end{figure}

Fig.~\ref{fig:res_turbo} shows the average and tail request latency for four configurations that combine enabling/disabling \hj{Turbo} and the $C1E$ and $C6$ C-states and highlight the effect of C-states on Turbo performance, compared to \tech's $C6A$ state with and without Turbo.

We make \textbf{three} key observations.
First, Fig.~\ref{fig:res_turbo}(a,c) show that the configuration with \hj{only} Turbo and $C6$ disabled (i.e., \texttt{NT\_No\_C6}) increases the average/tail latency performance by up to $4\%$/$31\%$ over the configuration with \hj{all of Turbo, $C6$, and} $C1E$ disabled (i.e., \texttt{NT\_No\_C6,No\_C1E}).
Second, comparing Figs.~\ref{fig:res_turbo}(c) and (d) \hj{shows} that enabling Turbo while disabling $C1E$ (i.e., \texttt{T\_No\_C6,No\_C1E}) does not improve performance over the same configuration with Turbo disabled (i.e., \texttt{NT\_No\_C6,No\_C1E}).
Third, Figs. \ref{fig:res_turbo}(b,d) show that with Turbo enabled, only disabling $C6$ (i.e., \texttt{T\_No\_C6}) has the same performance as additionally disabling $C1E$ (i.e., \texttt{T\_No\_C6,No\_C1E}).
The reason is that in the \texttt{T\_No\_C6} configuration, the transition overhead of $C1E$ on average/tail latency offsets any thermal capacitance gains and ensuing performance gains from Turbo.

We conclude that in a configuration where both $C6$ and $C1E$ are disabled while Turbo is enabled, large performance benefits can be obtained by \hg{enabling $C6A$ instead of $C1$}, i.e., \texttt{T\_C6A,No\_C6,No\_C1E}.
Doing so provides larger thermal capacitance to Turbo  compared to enabling $C1E$, and  reduces the long transition latency overhead of C-states (i.e., $C6$ and $C1E$). We illustrate the potential benefits of Turbo with \tech in Figs.~\ref{fig:res_turbo}(b,d) (dashed green line).      

\subsection{Analysis of Additional Workloads}\label{sec:other_workloads}
\begin{figure}[ht]
\centering
\includegraphics[trim=0.8cm 0.75cm 0.6cm 0.8cm, clip=true,width=1\linewidth,keepaspectratio]{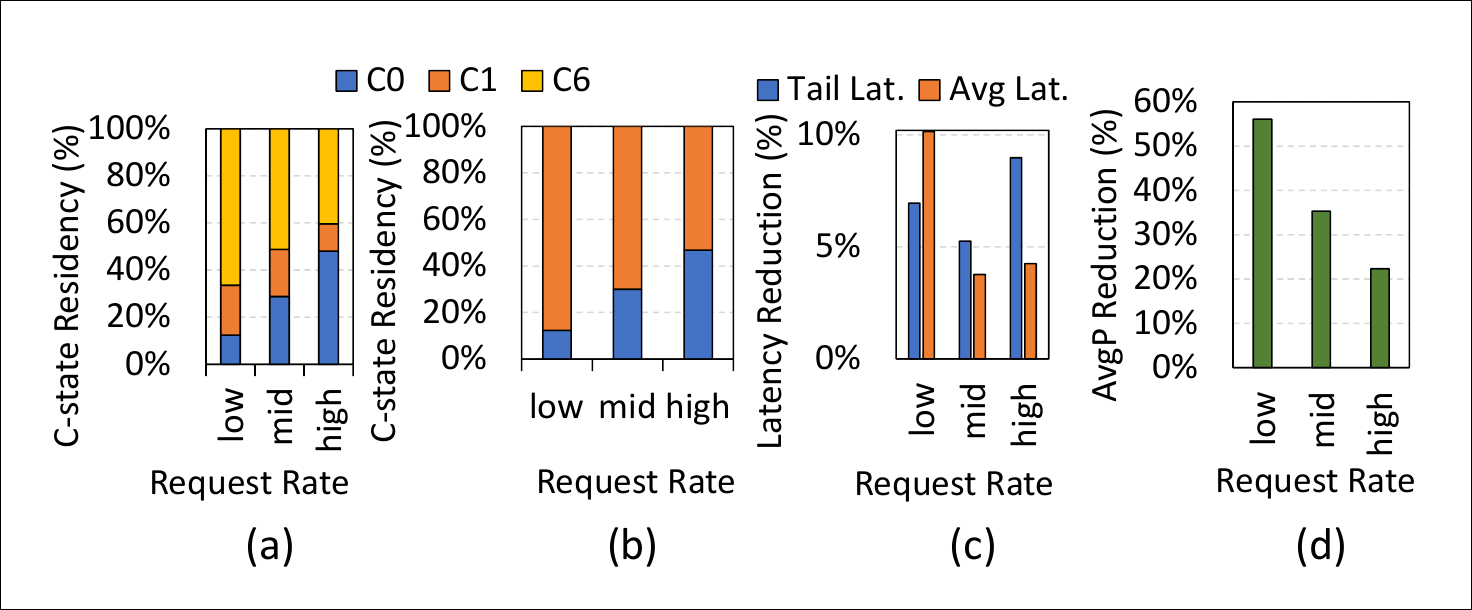}
\caption{Evaluation of {\tt MySQL} \cite{mysql} for low, mid and high request rates. (a) C-state residency of baseline and (b) with disabled $C6$ (c) Tail and average \hj{latency} \hjj{reduction} with $C6$ disabled (d) average power reduction with \tech's $C6A$ as compared to $C6$ disabled.}
\label{fig:res_mysql}
\end{figure}

Fig.~\ref{fig:res_mysql} shows the evaluation of {\tt MySQL} \cite{mysql}, a latency-critical workload, for three request rates (low, mid, and high). Fig. \ref{fig:res_mysql}(a) and \ref{fig:res_mysql}(b) show the C-state residency of the baseline configuration (P-states disabled,  $C1$ and $C6$ C-states enabled) and \hj{baseline} with $C6$ disabled\hj{, respectively}.

While for all request rates the baseline has ${\geq}40\%$ $C6$ residency, Fig. \ref{fig:res_mysql}(c) shows that the tail and average latency is significantly ($4\%$--$10\%$) improved when we disable $C6$.
Therefore, the recommended configuration is to disable  $C6$ (Fig. \ref{fig:res_mysql}(b)) to avoid its high transition latency. Fig. \ref{fig:res_mysql}(d) shows significant ($22\%$--$\%56$) average power reduction from \tech's $C6A$ as compared to a $C6$\hj{-}disabled configuration (i.e., $C1$ residency in Fig. \ref{fig:res_mysql}(b) mapped to $C6A$).

\begin{figure}[ht]
\centering
\includegraphics[trim=0.8cm 0.75cm 0.6cm 0.8cm, clip=true,width=1\linewidth,keepaspectratio]{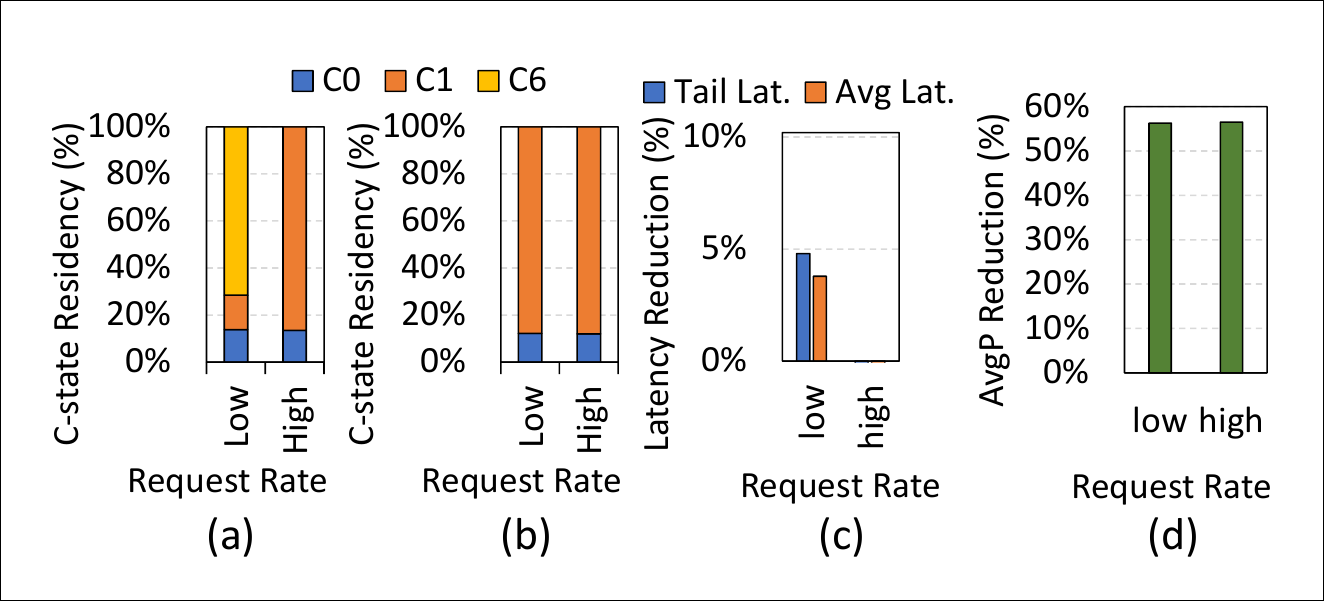}
\caption{Evaluation of {\tt Kafka} \cite{kreps2011kafka} for low and high request rates. (a)  C-state residency of baseline and (b) disabled $C6$ (c) Tail and average performance improvement with disabled $C6$ (d) average power reduction with \tech's $C6A$ as compared to $C6$ disabled.}
\label{fig:res_kafka}
\end{figure}

Fig. \ref{fig:res_kafka} shows the evaluation of {\tt Kafka} \cite{kreps2011kafka}\hj{,} another latency-critical workload for two request rates (low and high). Fig. \ref{fig:res_kafka}(a) and \ref{fig:res_kafka}(b) show the  C-state residency of the baseline configuration (P-state disabled,  $C1$ and $C6$ C-state enabled) and \hj{baseline with} $C6$ state \hj{disabled, respectively}.

At a low request rate\hj{,} the baseline has \hj{${>}60\%$} $C6$ residency. As shown in Fig. \ref{fig:res_kafka},  disabling \hj{the} $C6$ state improves the tail and average latency by $4\%$--5$\%$. The high request rate point does not show a performance improvement since at high rates {\tt Kafka} workload does not enter  \hj{the} $C6$ state. Fig. \ref{fig:res_kafka}(d) shows ${>}56\%$ average power reduction\hjj{, for both request rates,} by using \tech's $C6A$ \hjj{(i.e., $C1$ residency is mapped to $C6A$)} compared to having $C6$ disabled \hjj{(shown in Fig. \ref{fig:res_kafka}(b))}.  

We conclude that \tech with disabled P-states and enabled Turbo significantly improves core average power for \hj{the} {\tt MySQL} and {\tt Kafka} \hj{workloads} as compared to the baseline.

\subsection{Impact of High Snoop Traffic}\label{sec:snoop_impact}
To understand the impact of high snoop rate on \tech power savings, we analyze the power consumption difference between baseline and \tech while handling snoops. If a snoop arrives in baseline, then the system clock-ungates the L1/L2 subsystem (additional ${\sim}50mW$ to core $C1$ state) and handles the snoops. In \tech, L1/L2 exit the sleep mode and handle the snoops. The power difference, therefore, is mainly the L1/L2 exit from sleep mode (additional ${\sim}120mW$ to $C6A$). To calculate an upper bound on power savings opportunity of \tech compared to baseline with and without snoops, we assume a $100\%$ idle core where the $C1$ ($C6A$) state is the only state that is enabled (i.e., $R_{C1}=R_{C6A}=100\%$). If the core does not handle any snoop then \tech power savings are $(P_{C1}-P_{C6A})/P_{C1}=1.44-0.3)/1.44 = 79\%$. In case the core is handling snoops all the time during $C1$ ($C6A$) AW power savings are   $(1.49-0.470)/1.49 = 68\%$. Therefore, in the worst case we lose an $11\%$ power savings opportunity in case of high snoop traffic.

\subsection{Data Center Cost Savings Analysis}\label{sec:DC_cost_red}
AW enables significant power reduction even during short bursts of core idleness that are infeasible using existing core C-states. In a data-center context, \hj{all} else being equal, AW power savings translate to lower operational cost since less energy is consumed during  periods that cores enter the AW idle C-states as compared to residing \hj{in} the shallower legacy C-states. 
AW does not reduce cooling capital expenses since it does not reduce the TDP of a CPU\hj{, which} can be reached during times where a CPU has high utilization and cores do not enter idle states. As a result, a data-center that employs servers with CPUs supporting AW will need to provision for the worst-case cooling needs as with CPUs with only legacy C-states. Table \ref{tab:cost_savings} shows AW cost savings\footnote{The cost savings per server are calculated as (Average\_Baseline\_Power – Average\_AW\_Power) x Seconds\_in\_Year x Cost\_per\_Joule (assuming $\$0.125/KWh$ ~\cite{EnergyPrice}).} from operating a CPU during a year at different load levels, assuming the CPU  is running a Memcached workload. The savings range between \hj{$0.33$} to \hj{$0.59$} million dollars per year per 100K servers\hj{. These} savings grow proportionally to the data-center PUE~\cite{Jalili2021}. 
Besides CPU energy savings, \hj{AW's} lower idle power translates to lower time-averaged power and lower temperature that can extend a \hj{server's} lifetime (\hj{by slowing} down aging) and lower maintenance costs~\cite{Jalili2021}. A more detailed analysis of these aspects is beyond the scope of this work.

\begin{table}[h]
\centering
\caption{AW Yearly Cost Savings \hjj{(in \$M)} per 100K Servers}
\label{tab:cost_savings}
\resizebox{\linewidth}{!}{%
\begin{tabular}{|c||c|c|c|c|c|c|c|}
\hline
\multicolumn{1}{|c||}{\textbf{QPS}}        & \textbf{10K} & \textbf{50K} & \textbf{100K} & \textbf{200K} & \textbf{300K} & \textbf{400K} & \textbf{500K} \\ \hline \hline
\textbf{Savings (\$M/100K Servers)} & 0.33         & 0.59         & 0.58          & 0.53          & 0.47          & 0.41          & 0.34          \\ \hline
\end{tabular}%
}
\vspace{-14pt}
\end{table}

%% file: _08_relatedwork.tex
\section{Related Work}
\label{sec:related}
\jk{As far as we know, this is the first work to introduce a very low power processor core \hjj{power-}state \hjj{(i.e., deep idle C-state)} that provides both low transition latency and low power consumption.} While low server efficiency for latency-critical workloads has been studied before, previous work proposed management and scheduling techniques to mitigate the problem, rather than addressing it directly \jk{(i.e., with a fast yet low-power \hjj{processor core} power state)}.

\vspace{-1pt}
\noindent\hbb{\textbf{Modern Cloud Applications.}} Interactive latency-sensitive cloud applications are gradually shifting to a modular architecture based on loosely-coupled microservices to meet their software maintenance, scalability, and availability requirements~\mbox{\cite{B6, B7, B8}}.
However, the decoupled nature of microservices exacerbates the strict tail latency requirements of such applications. 

Recent work~\cite{B4} that characterizes large-scale deployments of microservices at Alibaba clusters shows that servicing a single user request may involve tens or even hundreds of microservices. 
With each microservice contributing a small amount of service time, together these can add up to a significant end-to-end service latency. 
Memcached appears in a significant fraction of \jk{microservices'} call graphs due to its ability to reduce the time to retrieve hot data from databases, making Memcached an important and latency critical component.
The same \jk{Alibaba} study also reveals that to meet the stringent tail latency requirements of microservices, Alibaba servers running latency-sensitive microservices typically operate at 10\% utilization. 
This follows previous reports from industry and academia that the utilization of servers running latency-sensitive applications is typically $5\%$--$20\%$~\mbox{\cite{lo2014towards,B1,B2,B3,B4,jalili2021cost}}. \hjj{Servers} at low utilization, however, \hjj{have} traditionally suffered from poor energy efficiency, partly because servers running latency-sensitive tasks at low utilization also disable deep C-states to avoid \hjj{the} \hjjj{tens or hundreds of microseconds} \hjj{penalty of}  transitioning out of such deep C-states~\mbox{\cite{cisco_cstates,dell_cstates}}. This emphasizes the need for deep C-states with low transition latency and high power savings.

\vspace{-1pt}
\noindent\textbf{Fine-grained, Latency-Aware DVFS Management.}
Besides C-states, the other major power-management feature of modern processors is dynamic voltage and frequency scaling (DVFS).
Previous work proposes fine-grained DVFS control to save power, while avoiding excessive latency degradation.
Rubik~\cite{kasture2015rubik} scales core frequency at \jk{sub-millisecond} scale based on a statistical performance model to save power, while \jk{meeting a} target tail latency.
Swan~\cite{zhou2020swan} extends this idea to computational sprinting\hjj{~\cite{raghavan2012computational,raghavan2013computational,raghavan2013utilizing,raghavan2013designing}} \hj{(e.g., Intel Turbo Boost \cite{rotem2013power,rotem2012power,rotem2011power})}: requests are initially served on a core operating at low frequency and, depending on the  load, Swan scales \jk{up} the frequency (including sprinting levels) to catch up and meet latency requirements.
NMAP~\cite{kang2021nmap} focuses on the network stack and leverages transitions between polling and interrupt \jk{modes} as a signal to drive DVFS management.
The new \CAgile state of \tech facilitates the effective use of idle states and \jk{could make} a simple race-to-halt approach \hjj{\cite{kim2015racing,albers2014race,awan2011enhanced,efraim2012energy,intel_skylake_2017,haj2018power}}  more attractive than complex DVFS management techniques.

\vspace{-1pt}
\noindent\textbf{Workload-Aware Idle State Management.}
Various proposals exist for techniques that profile incoming request streams and use that \jk{profile} information to improve power management.
SleepScale~\cite{liu2014sleepscale} is a runtime power management \jk{technique} that selects the most efficient C-state and DVFS setting for a given QoS constraint based on workload profiling information.
WASP~\cite{yao2017wasp} proposes a two-level power management framework;
the first level tries to steer bursty request streams to a subset of servers, such that other machines can leverage deeper, longer-latency idle states;
the second level adjusts local power management decisions based on workload characteristics such as job size, arrival pattern and 
utilization.
Similarly, CARB~\cite{zhan2016carb} packs requests into a subset of cores, while limiting latency degradation, so that the other cores have longer quiet times and can transition to deeper C-states.
The idea of packing requests \jk{into a} subset of active cores, to extend \jk{idle} periods on other cores is explored by work focusing on both C-state and DVFS management~\cite{chou2016dynsleep,asyabi2020peafowl,chou2019mudpm}.
\hgg{These proposals are orthogonal to \mbox{\tech}:} while \CAgile  can provide most of the benefits of a deep idle state at a much lower latency, advanced and workload-aware sleep management techniques can still bring extra power savings by enabling cores \hj{and/or system} to enter traditional deeper, higher-latency C-states \cite{meisner2009powernap,meisner2011power,pelley2009understanding,raghavendra2008no,antoniou2022agilepkgc}.
\hjj{Memory power management techniques \hjjj{\cite{delaluz2002scheduler,delaluz2001dram,diniz2007limiting,chang2017understanding,haj2020sysscale,david2011memory,deng2011memscale,chen2011predictive,deng2012coscale,felter2005performance,li2007cross,zhang2016maximizing,imes2018handing,deng2012multiscale}} have also been proposed to reduce system energy consumption and they are complementary to our work.}


%% file: _09_conclusion.tex
\section{Conclusion}
\jk{To our knowledge, \techL(\tech) is the first \hjj{deep idle} core power-state architecture that directly reduces \hjjj{the} transition latency to/from very low power states, with the goal of improving the energy efficiency of servers running latency-critical datacenter workloads.} 
Our evaluation reveals that \tech~\hjjj{realizes} power savings of up to $71\%$ per core \hjj{with ${<}1\%$  end-to-end} performance degradation. \jk{These results support the adoption of \tech in future datacenter server CPUs running \hjj{latency-critical applications}} and calls for further research
\jk{\hjj{into} lowering the latency of deep idle states.}
